\documentclass[journal]{IEEEtran}
\usepackage{amsthm}
\usepackage{amsmath,amsfonts}
\usepackage{algorithmic}
\usepackage{algorithm}
\usepackage{color}
\usepackage{array}
\usepackage[caption=false,font=normalsize,labelfont=sf,textfont=sf]{subfig}
\usepackage{textcomp}
\usepackage{stfloats}
\usepackage{url}
\usepackage{verbatim}
\usepackage{graphicx}
\usepackage{cite}
\usepackage[nottoc]{tocbibind}
\usepackage{tabularx}
\usepackage{pifont}
\usepackage[colorlinks=true,
            linkcolor=blue,    
            citecolor=blue,    
            urlcolor=blue]{hyperref}
\newcommand{\tabincell}[2]{\begin{tabular}{@{}#1@{}}#2\end{tabular}} 
\newtheorem{theorem}{Theorem}[section]
\newtheorem{lemma}[theorem]{Lemma}
\newtheorem{remark}{Remark}

\begin{document}

\title{Adaptive Subarray Segmentation: A New Paradigm of Spatial Non-Stationary Near-Field Channel Estimation for XL-MIMO Systems}

\author{
Shuhang Yang,~\IEEEmembership{Graduate Student Member,~IEEE}, Puguang An,  Peng Yang,~\IEEEmembership{Member,~IEEE}, 
 \\ Xianbin Cao,~\IEEEmembership{Senior Member,~IEEE}, Dapeng Oliver Wu,~\IEEEmembership{Fellow,~IEEE}, and Tony Q. S. Quek,~\IEEEmembership{Fellow,~IEEE}
 
\IEEEcompsocitemizethanks{\IEEEcompsocthanksitem S. Yang, P. An, P. Yang, and X. Cao are with the School of Electronic and Information Engineering, Beihang University, Beijing 100191, China.
\IEEEcompsocthanksitem D. O. Wu is with the Department of Computer Science, City University of Hong Kong, Kowloon, Hong Kong, China.
\IEEEcompsocthanksitem T. Q. S. Quek is with the Information Systems Technology and Design Pillar, Singapore University of Technology and Design, Singapore 487372, Singapore.
}
}

\maketitle

\begin{abstract}To address the complexities of spatial non-stationary (SnS) effects and spherical wave propagation in near-field channel estimation (CE) for extremely large-scale multiple-input multiple-output (XL-MIMO) systems, this paper proposes an SnS-aware CE framework based on adaptive subarray partitioning. We first investigate spherical wave propagation and various SnS characteristics and construct an SnS near-field channel model for XL-MIMO systems. Due to the limitations of uniform subarray patterns in capturing SnS, we analyze the adverse effects of the non-ideal array segmentation (over- and under-segmentation) on CE accuracy. To counter these issues, we develop a dynamic hybrid beamforming-assisted power-based subarray segmentation paradigm (DHBF-PSSP), which integrates power measurements with a dynamic hybrid beamforming structure to enable joint subarray partitioning and decoupling. A power-adaptive subarray segmentation (PASS) algorithm leverages the statistical properties of power profiles, while subarray decoupling is achieved via a subarray segmentation-based sampling method (SS-SM) under radio frequency (RF) chain constraints. For subarray CE, we propose a subarray segmentation-based assorted block sparse Bayesian learning algorithm under the multiple measurement vectors framework (SS-ABSBL-MMV).  This algorithm exploits angular-domain block sparsity under a discrete Fourier transform (DFT) codebook and inter-subcarrier structured sparsity. Simulation results confirm that the proposed framework outperforms existing methods in CE performance.
\end{abstract}


\section{Introduction}
\IEEEPARstart{T}{o} meet the ever-growing communication capacity demands of 6G networks, communication systems are shifting towards millimeter-wave (mmWave) and terahertz (THz) frequencies \cite{10884543,Cao2025Survey}. Due to the high path loss and short wavelength of high-frequency signals, the combination of extremely large-scale multiple-input multiple-output (XL-MIMO) and mmWave/THz technology enhances gain and spectral efficiency while maintaining a limited antenna size \cite{10500403,10132547}. However, with a dramatic increase in the number of antennas, the array's Rayleigh distance grows significantly, potentially placing the user equipment (UE) in the radiative near-field. In this region, electromagnetic wave propagation is better approximated by a spherical wave model rather than the conventional plane wave assumption \cite{10379539}. Additionally, the massive number of elements in XL-MIMO arrays allows different array regions to observe distinct scatterers, leading to the spatial non-stationary (SnS) effect \cite{9170651,8866736}. Channel state information (CSI) acquisition is crucial for optimizing wireless communication systems. However, due to the large number of elements and the impact of SnS in XL-MIMO systems, achieving high-accuracy CE with low pilot overhead remains a significant challenge. 

\subsection{Prior Works}
\par \textbf{Near-field channel estimation:} Under the far-field plane-wave assumption, exploiting the angular-domain sparsity of the mmWave/THz channel, numerous compressive sensing (CS) algorithms have been employed for CE with low pilot overhead. These include the well-known orthogonal matching pursuit (OMP) algorithm \cite{7458188}, sparse Bayesian learning (SBL) \cite{6826590}, variational Bayesian inference (VBI) algorithm \cite{8038934}, and approximate message passing (AMP) algorithm \cite{8630098}. To address the on-grid mismatch issue, \cite{8370683} and \cite{8432470} proposed the iterative reweighted super-resolution compressive sensing approach and the atomic norm minimization-based gridless compressive sensing algorithm, respectively.

\par Under the near-field spherical-wave assumption, the inherent nonlinear coupling between distance and angle in the steering vector invalidates conventional far-field CE methods. To address this, the authors in \cite{9693928} introduced a polar-domain codebook that jointly samples distance and angle, along with an on/off-grid simultaneous OMP (SOMP) algorithm. This two-dimensional (2D) sampling generates a large-scale sparsifying dictionary, leading to high computational complexity. Additionally, the strong correlation among near-field steering vectors degrades the restricted isometry property (RIP), further limiting CE performance. Building on this, \cite{10273424} proposed a distance-parameterized angular-domain sparse model, where user distance is incorporated as an unknown parameter in the dictionary. Unlike distance–angle representations, \cite{10819473,10637261} approximate spherical waves via a Fourier plane-wave expansion and employ the continuous Fourier transform (CFT) to construct a wavenumber codebook, thereby mitigating grid mismatch. Separately, \cite{10845870} designed a discrete prolate spheroidal sequence (DPSS)-based codebook and a two-step CE scheme, though it requires prior rough UE location estimation. Differing from these approaches, \cite{10638078} explored the block sparsity shared between near-field and far-field angular domains and proposed a complex simultaneous logit-weighted block OMP (CSLW-BOMP) algorithm to exploit this structural characteristic. 

\par \textbf{XL-MIMO channel estimation affected by SnS:} Existing XL-MIMO CE methods that account for SnS can be broadly classified into two categories. The first category is subarray-based methods, in which the full array is partitioned into equal-sized subarrays and channel estimation is performed on each subarray \cite{8949454,10373799, 10631699,10345492,10815057}. To address the subarray coupling problem under a limited number of radio frequency (RF) chains, \cite{10373799} proposed a group time block code (GTBC)–assisted subarray decoupling scheme for a fully-connected hybrid beamforming (HBF) architecture. GTBC is simple to implement but requires a large number of pilots for decoupling. Building on this idea, \cite{xu2025exploitingdynamicsparsitynearfield} adapted the HBF architecture to a partially-connected form and decoupled element-level received signals. Subarray-based approaches are attractive due to their simplicity and low complexity. However, equal-sized partitioning cannot guarantee that the visibility region (VR) birth-death will align with predefined subarray boundaries. The second category explicitly models the VR birth–death process in the array domain. For example, \cite{9547795} modeled VR birth-death as a Markov chain and exploited shared VRs across different paths to design structured priors. \cite{10715712,10509715} adopted a two-stage architecture consisting of VR detection followed by CE: \cite{10715712} employed a 2-D Markov prior to capture VRs in a uniform planar array (UPA) and then performed CE using an inverse-free variational Bayesian inference (IF-VBI) algorithm. While \cite{10509715} estimated the wavenumber-domain channel via belief-based OMP (BB-OMP). To reduce the complexity of jointly estimating VRs for multiple paths, \cite{10153711,10780971} first separated the channel into sub-channels using techniques such as the multiple signal classification (MUSIC) algorithm, and then performed VR detection and channel estimation on each sub-channel. Although these VR-prior methods offer high flexibility, their VR detection modules often require message passing across the array dimension and matrix inversions, resulting in higher computational complexity. Unlike VR-prior modeling, the approach in \cite{9777939} identified the user’s VR by exploiting the statistical characteristics of the received power across antenna elements. While this statistic depends on the accuracy of channel estimation and deteriorates in low signal-to-noise ratio (SNR) conditions, it provides a new perspective.

\subsection{Our Contributions}
Considering the issues in existing research about near-field CE under SnS, this paper proposes an adaptive subarray partitioning approach. The proposed method preserves the low complexity of subarray-based schemes while addressing the incomplete alignment between uniform subarray patterns and SnS. Moreover, it exploits block sparsity in the near-field angular domain and develops an efficient subarray-level CE algorithm. The main contributions are summarized as follows.

\begin{itemize}
\item{We first investigate the near-field characteristics in XL-MIMO systems, encompassing spherical wave propagation and SnS. Building upon these characteristics, we innovatively extend the conventional SnS near-field channel model by integrating non-ideal propagation paths encountered in obstructed line-of-sight (OLoS) scenarios and apply VR-based weighting masks to capture various SnS effects.}
\item{This paper introduces the concepts of over- and under-segmentation as non-ideal subarray partitioning modes and provides a theoretical analysis of their detrimental impact on channel estimation. To address these challenges, a novel framework employing low-cost power measurements combined with a dynamic hybrid beamforming (DHBF) architecture is proposed for joint subarray partitioning and decoupling. Specifically, the power-adaptive subarray segmentation algorithm (PASS) is developed using the statistical properties of the power profiles. Under limited RF chains, the DHBF architecture is combined with an RF chain allocation strategy and the subarray-segmentation-based sampling method (SS-SM) to achieve effective subarray decoupling. }
\item{Based on the subarrays obtained from the proposed partitioning, we develop a subarray-segmentation-based assorted block sparse Bayesian learning method under the multiple-measurement-vectors framework (SS-ABSBL-MMV). SS-ABSBL-MMV exploits block sparsity in the angular domain of near-field channel and structured sparsity across subcarriers. 
It performs channel estimation per subarray and is developed in both on-grid and off-grid versions. 
Finally, we derive the Bayesian Cramér–Rao bound (BCRB) to quantify the effectiveness of the proposed algorithms.}
\item{Simulation results demonstrate that the proposed DHBF-PSSP framework effectively supports SnS channel estimation while reducing pilot overhead. Furthermore, the proposed SS-(OG)-ABSBL-MMV algorithm still outperforms other existing algorithms in CE performance, even when using a DFT codebook with a significantly smaller number of atoms compared to polar-domain codebooks.}
\end{itemize}

\subsection{Organization and Notations}
\textit{Organization:} The rest of this paper is organized as follows. In Section \ref{sec2}, we present the SnS near-field channel model for XL-MIMO systems. In Section \ref{sec3}, we derive and elaborate on the importance of precise array partitioning, proposing the DHBF-PSSP and validating its effectiveness. In Section \ref{sec4}, we provide a detailed description of the SS-ABSBL-MMV algorithm. In Section \ref{sec5}, simulation results are presented and analyzed, followed by the conclusion in Section \ref{sec6}.
\par \textit{Notations:} Lowercase and uppercase bold letters denote vectors and matrices, respectively. Let $(\cdot)^{-1}$, $(\cdot)^\text{T}$, $(\cdot)^\text{H}$, $\|\cdot\|$, $\text{vec}(\cdot)$, $\text{det}(\cdot)$, $\text{tr}(\cdot)$, represent the inverse, transpose, conjugate transpose, ${\ell}_2$-norm, vectorization, determinant and trace, respectively. $\text{diag}(\cdot)$ is used to extract diagonal elements from a matrix or to construct a matrix with a vector as its diagonal elements. $\otimes$ is the Kronecker product operator and $\odot$ means the Hadamard product operator. $\Re\{\cdot\}$ denote the real part of the complex argument. $\mathbf{I}_N$ is the $N \times N$ dimensional identity matrix, and $\mathbf{1}_{M\times N}$ is the $M\times N$ dimensional all-one matrix. $\mathbf{e}_N$ is the standard basis vector.   For a set, $\lvert \cdot \rvert$ denotes the cardinality of the set. $\mathcal{CN}(\mathbf{x}; \boldsymbol{\mu}, \boldsymbol{\Sigma})$ denotes the complex Gaussian distribution with mean $\boldsymbol{\mu}$ and covariance $\boldsymbol{\Sigma}$.

\section{SnS Near-Field Channel Model} \label{sec2}
\subsection{Near-field Spherical Wave Channel Model}

\par XL-MIMO systems employ large-scale antenna arrays with $N$ elements at the base station (BS),  where the increased aperture quadratically extends Rayleigh distances. This likely positions scatters/UEs in the BS's radiating near-field, requiring spherical wave models for channel characterization: 
\begin{equation}
\label{near_field_channel}
\mathbf{h}=\sum_{l=1}^L g_l e^{-j \frac{2 \pi f}{c} r_l} \cdot \mathbf{b}\left(r_l, \theta_l\right), 
\end{equation}
\textcolor{red}{where $\mathbf{h}\in\mathbb{C}^N$ is the near-field spherical-wave channel model.} $L$ denotes the number of propagation paths, $g_l$ represents the complex channel gain of the $l$-th path. $f$ and $c$ denote the carrier frequency and the speed of light. $r_l$ and $\theta_l$ represent the distance and angle between the $l$-th scatterer/UE and the reference position (located at the array’s geometric center), respectively. \textcolor{red}{The near-field steering vector $\mathbf{b}\left(r_l, \theta_l\right)\in\mathbb{C}^N$ can be expressed as:}
\begin{equation}
\label{near_field_steering_vector}
\mathbf{b}\left(r_l, \theta_l\right)=\frac{1}{\sqrt{N}}\left[e^{-j k_w\left(r_l^{(1)}-r_l\right)}, \ldots, e^{-j k_w\left(r_l^{(N)}-r_l\right)}\right]^{\mathrm{T}},
\end{equation}
where $k_w=2 \pi f/c$ denotes the wavenumber. \textcolor{red}{And $r_l^{(n)}=\sqrt{r_l^2+(\Delta_nd)^2-2\Delta_ndr_l\sin\theta_l}$ denotes the distance between the $l$-th scatterer/UE and the $n$-th antenna element. Based on the geometric relationship and Taylor expansion\cite{9693928}, it can be approximated as:}
\begin{equation}
\label{near_field_distance}
r_l^{(n)} \approx r_l - \Delta_n d \sin \theta_l + \frac{\left(\Delta_n d \cos \theta_l\right)^2}{2 r_l},
\end{equation}
where $d$ represents the antenna spacing, which is typically set as half-wavelength spacing. And index $\Delta_n=\frac{2n - N - 1}{2}, n = 1,\ldots,N$.

\subsection{SnS and its Impact}
\par The SnS arises in XL-MIMO systems with array apertures spanning hundreds of wavelengths. It can be summarized as the following three aspects:
\begin{itemize}
    \item{Channel measurement results in \cite{10416965} demonstrate that within the near-field region of XL-MIMO systems, the received power distribution across the antenna array exhibits non-uniform characteristics due to the distinct propagation distances between the scatters/UEs and individual antenna elements\cite{8736783}.}
    \item{The array's extensive spatial coverage results in partial blocking of some paths, where scatters/UEs can only access a subset of antennas, defined as the VR at the base station (BS-VR), where signal power is considerably higher \cite{7062910}. 
    Meanwhile, SnS causes different antenna elements to observe different visible paths, naturally partitioning the array into multiple subarrays. Here, a subarray is defined as a contiguous segment over which the set of visible paths remains unchanged, i.e., a stationary interval (SI).}
    \item{In obstructed LoS (OLoS) scenarios, the direct path is partially blocked and diffracted waves from obstacle edges generate non-ideal components, which induce element-dependent power fluctuations across the array and thus cause spatial variations within the BS-VR \cite{10416965,10694152}.}
\end{itemize}
\begin{figure}[!t]
\centering
\includegraphics[width=3.5in]{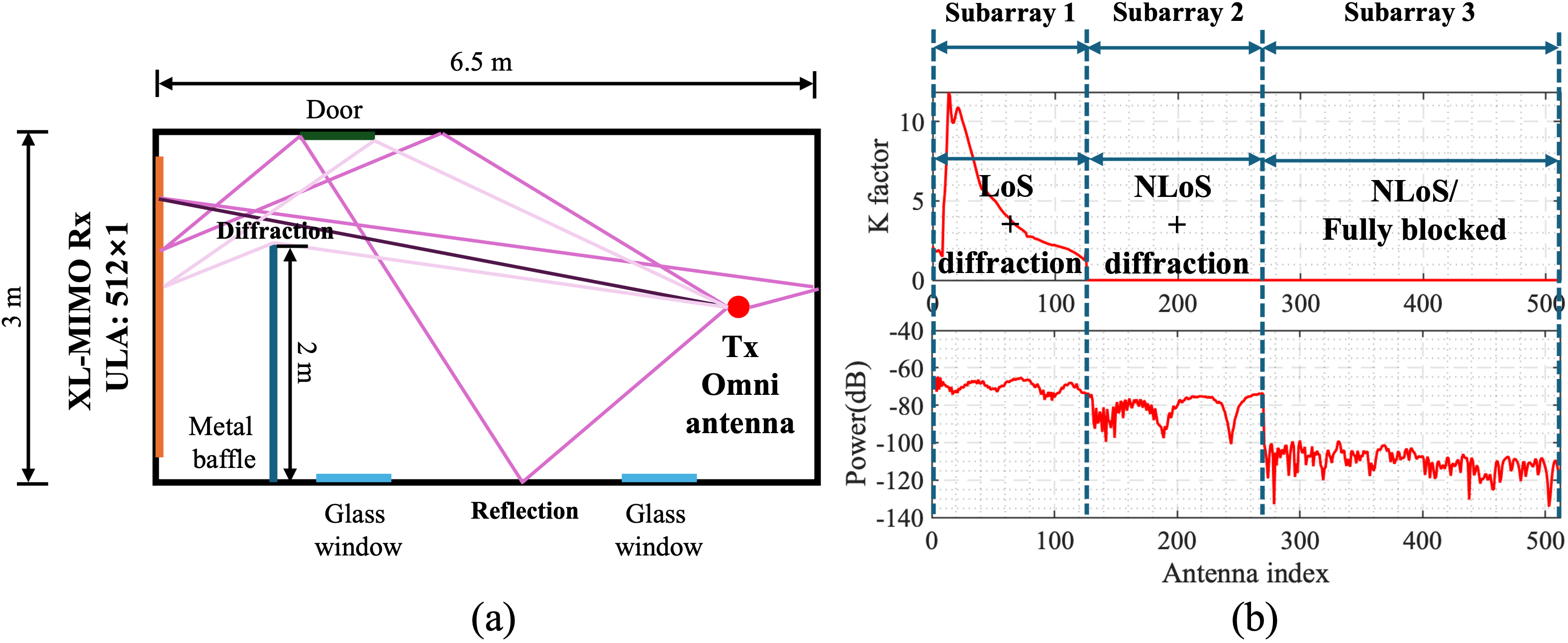}
\caption{SnS near-field channel for XL-MIMO system based on ray tracing with Remcom Wireless Insite: the transmitter and receiver utilize omnidirectional antennas and employ a 28 GHz narrow-band sinusoidal waveform.} 
\label{fig_1}
\end{figure}

\par The ray-tracing result shown in Fig. \ref{fig_1} validates the above properties. Distinct sets of visible paths create varying subarrays at the XL-MIMO array side. The subarray with strong LoS links (large Rician factor) exhibits substantially higher power, though power fluctuations due to non-ideal propagation effects remain observable. Inherently low-power NLoS paths exhibit more significant diffraction-induced fluctuations, while purely NLoS or fully blocked regions have near-zero power.

\par Considering these factors, we propose an enhanced channel model that extends the XL-MIMO framework to support SnS scenario. The modified channel model is expressed as:
\begin{equation}
\label{SnS_LOS_XLMIMO}
\mathbf{h}=\sum_{l=1}^L g_l e^{-j \frac{2 \pi f}{c} r_l} \cdot \mathbf{b}\left(r_l, \theta_l\right)\odot\mathbf{s}_l.
\end{equation}
\par The SnS properties are characterized through element-wise weighted masking of the channel matrix via vector \textcolor{red}{$\mathbf{s}_l \in \mathbb{R}^N_{\ge
0}$}, with its elements defined as \cite{9940939}:
\begin{equation}
\label{SnS_s}
s_l^{(n)}=\left\{\begin{array}{l}
R_{l,\text{IP}}^{(n)}  , n \in \mathrm{BS}-\mathrm{VR} , l\in\text {ideal propagation} \\
R_{l,\text{NIP}}^{(n)}  , n \in \mathrm{BS}-\mathrm{VR} ,  l\in\text {non-ideal propagation} \\
0  ,n \notin \mathrm{BS}-\mathrm{VR}
\end{array}\right..
\end{equation}

\par In the above expression, for the $l$-th path under ideal (non-diffracted) propagation, power variations within a BS-VR depend only on the distances between scatterers/UEs and antenna elements. Specifically, by normalizing the received power of the $l$-th path at a reference antenna, the mask at the $n$-th antenna is directly given by:
\begin{equation}
    R_{l,\text{IP}}^{(n)}=\sqrt{\frac{\lambda^2/(4\pi r_l^{(n)})^2}{\lambda^2/(4\pi r_l)^2}}=\frac{r_l}{r_l^{(n)}},
\end{equation}
where $\lambda$ is the wavelength of the signal.\footnote{\textcolor{red}{Under full visibility and ideal propagation, the proposed model reduces to the non-uniform spherical wave model (NUSW).}}  
\par In contrast, for non-ideal propagation, diffraction effects must be considered in addition to the above factors. To balance physical fidelity and computational efficiency, we adopt the single knife-edge diffraction model to construct the non-ideal propagation mask. This model assumes a sharp-edged obstacle between the transmitter and receiver \cite{itur}, and the diffraction gain (or loss) at the $n$-th antenna is expressed as:
\begin{equation}
    A_n=\frac{1}{4}\left[\left(1-\mathcal{C}(\nu_n)-\mathcal{S}(\nu_n)\right)^2+\left(\mathcal{C}(\nu_n)-\mathcal{S}(\nu_n)\right)^2\right],
\end{equation}
where $\mathcal{S}(\cdot)$ and $\mathcal{C}(\cdot)$ denote the Fresnel Sine and Cosine integral, respectively. Here, the Fresnel-Kirchhoff diffraction parameter $\nu_n$ for the $n$-th antenna is given by
\begin{equation}
    \nu_n =h_n\sqrt{\frac{2(d_{1,n}+d_{2,n})}{\lambda d_{1,n}d_{2,n}}},
\end{equation}
where $h_n$ is the height of the sharp obstacle above/under LoS between the $n$-th antenna of BS and UE \footnote{When the obstacle edge is above the LoS path between the $n$-th antenna of BS and UE, $h_n$ is positive, and it is negative when under.}. $d_{1,n}$ and $d_{2,n}$ are the distances from the $n$-th antenna of BS and UE to the obstacle, respectively. 
\par Let the parameters corresponding to the link between the reference position on the BS side and the UE be defined as $h_{ref}$, $d_{1,ref}$, and $d_{2,ref}$. By using the geometric relationship, the parameters of each antenna element can be obtained:
\begin{equation}
\begin{aligned}
    h_n&= \frac{h_{ref}q_n+d_{2,ref}\delta_n\cos{\theta}}{\sqrt{q_n^2+(\delta_n\cos{\theta})^2}},\\
        d_{2,n}&=\frac{d_{2,ref}q_n-h_{ref}\delta_n\cos{\theta}}{\sqrt{q_n^2+(\delta_n\cos{\theta})^2}},\\
        d_{1,n}& \approx (d_{1,ref}+d_{2,ref})+\delta_n \sin \theta
        \\ &+ \frac{\left(\delta_n \cos \theta\right)^2}{2(d_{1,ref}+d_{2,ref})}-d_{2,n},
\end{aligned}
\end{equation}
where $\theta$ represents the angle between the UE and the reference position. Other intermediate parameters are given as follows:
\begin{equation}
    \begin{aligned}
        \delta_n &= \frac{N-2n-1}{2}d,\ n=1\cdots N,\\
        q_n &= d_{1,ref}+d_{2,ref}+\delta_n\sin{\theta} .
    \end{aligned}
\end{equation}

\begin{figure}[!t]
\centering
\includegraphics[width=3in]{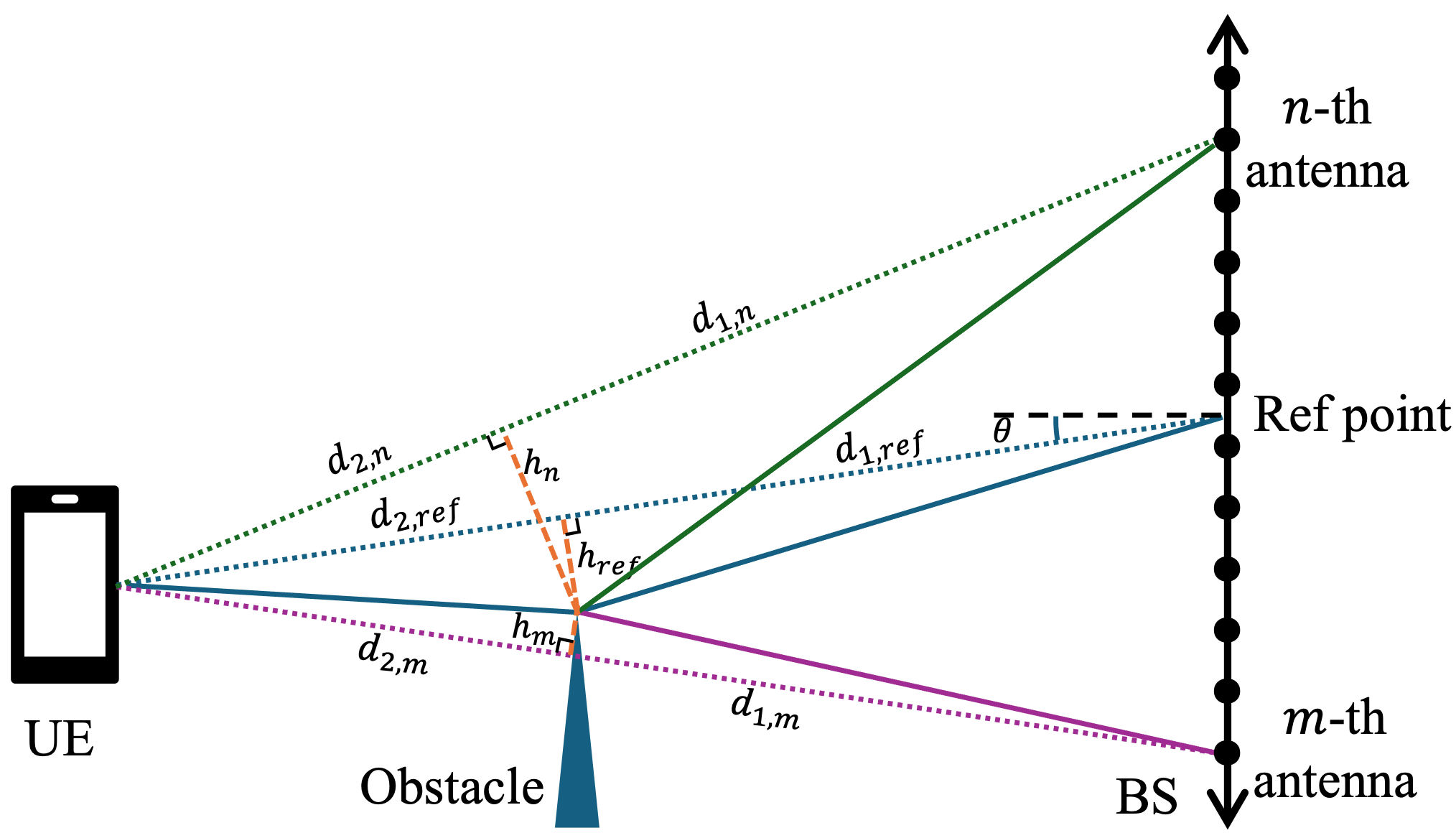}
\caption{Geometric schematic diagram of the single knife-edge diffraction model.}
\label{diffraction}
\end{figure}

\par Fig. \ref{diffraction} illustrates the geometry of the single-edge knife diffraction model. Notably, the diffraction parameter $\nu_n$ is a dimensionless measure of the obstruction level: larger positive $\nu_n$ implies stronger attenuation due to blockage, whereas $\nu_n<0$ indicates weaker attenuation but observable diffraction-induced fluctuations.

\par Taking both spherical-wave propagation and diffraction into account, the mask for a non-ideal propagation path is given by:
\begin{equation}
    R_{l,\text{NIP}}^{(n)}=\frac{r_l}{r_l^{(n)}}\left[t_d\cdot(\sqrt{A_n}-1)+1\right],
\end{equation}
where $t_d$ controls the diffraction intensity and must satisfy $t_d<1/(1-\text{min}(A_n))$ to ensure the mask remains positive 

\begin{remark}
    The BS-VR birth–death behavior of each path is modeled separately according to the path type. For ideal propagation paths, the BS-VR birth–death process is represented as a block-level first-order Markov chain \cite{10715712,9718019}. Specifically, the $N$-element aperture is divided into $B_\text{SI}=\lceil N/\mathrm{SI}_{\min}\rceil$ contiguous blocks of length $\mathrm{SI}_{\min}$, indexed by $b=1,\dots,B_\text{SI}$. For each path $l$, we define a binary block state $z_{b,l}\in\{0,1\}$, where $z_{b,l}=1$ indicates that block $b$ lies within the BS-VR of path $l$. The evolution of the sequence $\{z_{b,l}\}_{b=1}^{B_\text{SI}}$ is governed by the first-order Markov transition matrix. For non-ideal paths, diffraction-induced power variations depend on both obstacle geometry and wavelength. These effects are captured element-wise by the single knife-edge diffraction model through the diffraction gain $A_n$. Therefore, the BS-VR is determined by thresholding the per-element power rather than by a stochastic birth–death generator \cite{10999186}.
\end{remark}

\section{Subarray Segmentation: Theoretical Importance and Innovative Paradigm} \label{sec3}
Conventional channel estimation methods fail to capture SnS, resulting in degraded estimation accuracy. Existing subarray-based approaches preset uniform subarray sizes, which cannot guarantee that SnS birth–death will coincide with predefined subarray boundaries. To address these limitations, we first demonstrate the adverse impact of this phenomenon. Then, we propose an adaptive subarray segmentation applied before CE and perform per-subarray estimation to capture SnS.
\subsection{The Importance of Subarray Segmentation}
\begin{figure}[!t]
\centering
\includegraphics[width=2.7in]{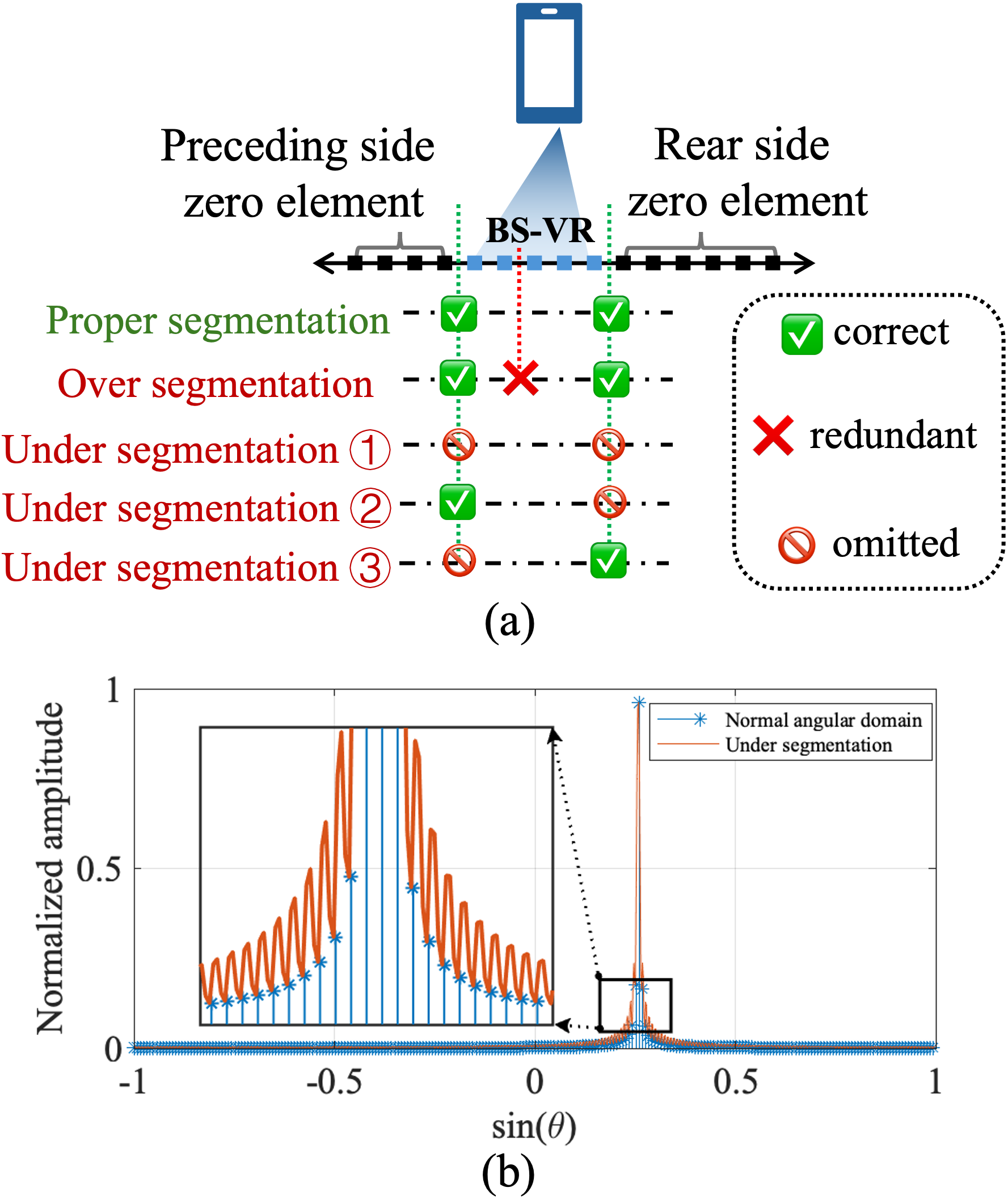}
\caption{Subarray segmentation, (a) {Three segmentation patterns}, (b) Effect of under-segmentation on angular-domain channel.}
\label{fig_2}
\end{figure}
To illustrate, we will discuss two scenarios of non-ideal subarray segmentation for the sake of brevity: \textit{over-segmentation} and \textit{under-segmentation}.
\subsubsection{Over-segmentation}{Over-segmentation refers to the redundant segmentation placed at antenna positions where no SnS birth–death actually occurs, as Fig. \ref{fig_2}(a). These redundant segments reduce the number of elements per subarray, thereby directly degrading the angular resolution. Because compressed-sensing channel estimation needs adequate angular resolution to separate sparse components, over-segmentation degrades its accuracy.}
\subsubsection{Under-segmentation}{Under-segmentation mixes BS-VR elements and neighboring non-BS-VR elements into the same subarray, altering the subarray’s angular representation and undermining sparsity.}
\par {To illustrate the hazards of under-segmentation, consider a single path whose BS-VR spans $Q$ elements, as Fig. \ref{fig_2}(a). Let $P_0$ and $R_0$ denote the number of non-BS-VR elements on the preceding and rear sides, respectively. The spatial channel can be written as:}
\begin{equation} {
	    	\mathbf{h}(n)=
            \begin{cases}
            0, & n=0,\ldots,P_0-1,\\
            \mathbf{h}_{\mathrm{VR}}(n), & n=P_0,\ldots,P_0+Q-1,\\
            0, & n=P_0+Q,\ldots,P_0+Q+R_0-1
            \end{cases},}
\end{equation}
where $\mathbf{h}_{\mathrm{VR}}$ denotes the spatial channel within the BS-VR.
\par The correct segmentation would split at $n=P_0$ and $n=P_0+Q$, producing three subarrays. Under segmentation in mode \ding{172} in Fig. \ref{fig_2}(a) ignores all correct segmentation points. The Fourier transform of $\mathbf{h}(n)$ gives its angular representation: 

\begin{equation}
\label{overseg_eq}
\begin{split} 
\mathbf{h}_a(k) =& \sum_{n=0}^{S-1} \mathbf{h}(n) e^{-j \frac{2 \pi n}{S} k} \\
=& e^{-j \frac{2 \pi k}{S} P_0}\sum_{u=0}^{Q-1} \left[\frac{1}{Q}\sum_{m=0}^{Q-1}\mathbf{h}_{a,\text{VR}}(m)e^{j\frac{2\pi}{Q}mu}\right] e^{-j \frac{2 \pi k}{S} u} \\
=& e^{-j \frac{2 \pi k}{S} P_0} \sum_{m=0}^{Q-1}\frac{\mathbf{h}_{a,\text{VR}}(m)}{Q}\frac{1-e^{-j2\pi\frac{kQ}{S}}}{1-e^{j\frac{2\pi}{Q}(m-\frac{k}{S/Q})}}, \\
&k = 0,\ldots,S-1, 
\end{split}
\end{equation}
where $S=P_0+Q+R_0$ is the total number of elements, and $\mathbf{h}_{a,\mathrm{VR}}$ denotes the angular-domain representation of $\mathbf{h}_{\mathrm{VR}}$. 
\par {Clearly, Eq. (\ref{overseg_eq}) shows that $\mathbf{h}_a$ is an interpolation of $\mathbf{h}_{a,\text{VR}}$, as shown in Fig. \ref{fig_2}(b). Both mode \ding{173} and mode \ding{174} can be viewed as special cases of mode \ding{172}. In practice, non-ideal segmentation might be a combination of these modes, but its primary effect on the angular domain channel is interpolation.} This is similar to how zero-padding in the time domain leads to frequency domain interpolation in digital signal processing. Although zero-padding can enhance spectral resolution and reduce the picket-fence effect in time-frequency transforms \cite{blackledge2006digital}, such interpolation disrupts sparsity and leads to misidentification of the support set in the angular domain, severely degrading CS performance. Thus, precise identification of SnS is critical for near-field CE.

\subsection{DHBF-PSSP Overview}

\begin{figure}[!t]
\centering
\includegraphics[width=3.1in]{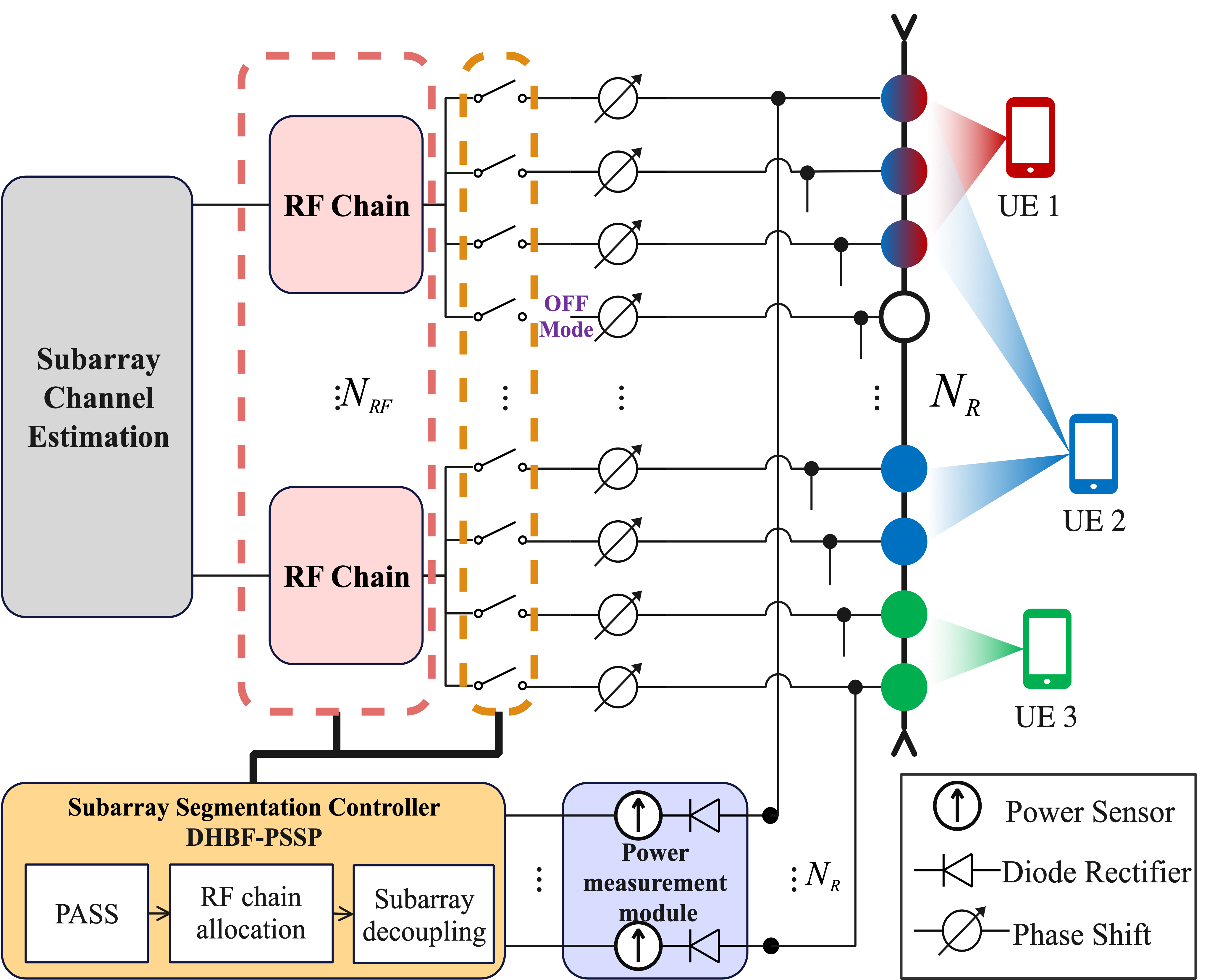}
\caption{{Diagram of DHBF-PSSP.} }
\label{fig_3}
\end{figure}

Before CE, due to the limited features that can be extracted from the signal, signal power, as a readily measurable metric, serves as a key subarray partitioning criterion. 
Based on this, we propose an innovative subarray segmentation paradigm, termed the DHBF-assisted power-based subarray segmentation paradigm (DHBF-PSSP), as illustrated in Fig. \ref{fig_3}.
\begin{itemize}
\item{\textbf{Power Measurement and Subarray Partitioning}: {A low-power(milliwatt‑level), cost-effective sensor at the antenna feed measures per-element received power and is already in practice \cite{10042005, MaHongGaoJingLiuBaiChengCui+2020+3271+3278}. Using power measurements from this hardware, we developed a power-adaptive subarray segmentation algorithm (PASS) that determines each subarray’s extents by detecting edge points, as detailed in Section \ref{PASS_sec}.}}
\item{\textbf{DHBF-Assisted Subarray Decoupling}: {Based on the subarray extents obtained with PASS, the DHBF architecture assigns RF chains to subarrays to perform CE. In practice, the number of RF chains is often smaller than the number of subarrays, so a single RF chain may serve multiple subarrays, inducing received signal coupling that prevents CE \cite{10373799}. We propose a subarray-segmentation-based sampling method (SS-SM) that decouples received signals among subarrays sharing an RF chain, thereby providing uncoupled observations for subsequent subarray channel estimation.}}
\item{\textbf{RF Chain Resource Allocation Strategy}: {Given a fixed number of RF chains and an uncertain number of subarrays, RF-chain allocation must be addressed. If the number of subarrays is less than or equal to the number of RF chains, each subarray is assigned at least one dedicated RF chain. Otherwise, an allocation strategy that accounts for the relative counts and the properties of SS-SM is required, as detailed in Section \ref{DHBF_sec}.}}
\end{itemize}

\subsection{Power-based Subarray Segmentation Algorithm}
\label{PASS_sec}
Inter-subarray discrepancies in multipath number and power induce pronounced statistical divergence in aggregate power. Within a subarray, the spherical wave and non-ideal propagation of some paths can lead to power fluctuations. Such fluctuations, when combined with noise, may result in the under-segmentation between subarrays or the over-segmentation within a subarray.
\par Consider a communication system where $M$ subcarriers serve $K$ single-antenna UEs. Based on the channel model in Eq. \eqref{SnS_LOS_XLMIMO}, where each UE's channel comprises $L$ paths, and simplifying the subscripts, the power received at the $n$-th antenna array element can be expressed as:
\begin{equation}
p_n=\left|\sum_{k=1}^K \sum_{l=1}^L\sum_{m=1}^M\tilde{g}_{k,l}^{(n)} e^{-j \frac{2 \pi f_m}{c} r_{k,l}} b^{(n)}\left(r_{k,l}, \theta_{k,l}\right)\right|^2,
\end{equation}
where the equivalent complex channel gain $\tilde{g}_{k,l}^{(n)}=g_{k, l} s_{k,l}^{(n)}$, and $b^{(n)}\left(r_{k,l}, \theta_{k,l}\right)$ denotes the $n$-th element of the steering vector $\mathbf{b}\left(r_{k, l}, \theta_{k, l}\right)$.

\par To demonstrate that the statistical properties of the power of distinct incident waves can be leveraged for subarray segmentation, we present the following lemma.
\begin{lemma}\label{lem1}
	Under an ideal propagation scenario ($s_{k,l}^{(n)}=1, \forall n,k,l$) with a uniform linear array (ULA) comprising $N$ antennas and $K$ visible UEs, the channel gain \(g_{k,l}\sim\mathcal{CN}(0,\sigma^2)\). $g_{k,l}, \theta_{k,l}$, and $r_{k,l}$ are all mutually independent. The mean and variance of received power at an arbitrary array element are given by
    \begin{equation}
        \mathbb{E}[p_n] = \frac{MKL\sigma^2}{N}, \quad \mathrm{Var}[p_n] = \frac{M^2K^2L^2\sigma^4}{N^2}.
    \end{equation}
\end{lemma} 

\begin{IEEEproof}[Proof]
Let $h_{k,l,m}^{(n)}=g_{k,l} \cdot e^{-j \frac{2 \pi f_m}{c} r_{k,l}} \cdot b^{(n)}\left(r_{k,l}, \theta_{k,l}\right)$ denote the $n$-th element of the channel matrix $\mathbf{h}_{k,l,m}$ for the $l$-th path of UE $k$ at the $m$-th subcarrier frequency.  
Since \(g_{k,l}\sim\mathcal{CN}(0,\sigma^2)\) are statistically independent of $\theta_{k,l}$, and $r_{k,l}$. By the rotational invariance of circularly symmetric complex Gaussian variables \cite{tse2005fundamentals}: 
\[
h_{k,l,m}^{(n)}\sim\mathcal{CN}(0,\frac{\sigma^2}{N}).
\]
\par Furthermore, by the additivity of independent Gaussian random variables: 
\[
h_\text{all} = \sum_{m=1}^M \sum_{l=1}^L \sum_{k=1}^K h_{k,l,m}^{(n)} \; \sim \; \mathcal{CN}(0,\sigma^2_\text{all}), \;\sigma^2_\text{all}=\frac{MKL\sigma^2}{N}
\]
\par Since \(p_n=\left|h_\text{all}\right|^2\), \(p_n\) follows an exponential distribution with rate parameter \(1/\sigma^2_\text{all}\). Using the properties of the exponential distribution, we obtain: 
\[
\mathbb{E}\left[p_n\right] = \frac{MKL \sigma^2}{N},
\quad
\mathrm{Var}\left[p_n\right] = \frac{M^2K^2L^2 \sigma^4}{N^2}.
\]
\end{IEEEproof}
Lemma \ref{lem1} establishes that incoming wave count and power govern subarray-level received power statistics. While derived under large-sample assumptions, this principle retains operational validity for subarray partitioning in small-$KL$ regimes.

\par Inspired by sequential change-point detection \cite{8736886}, we reformulate array partitioning as a change-point detection problem based on the array power sequence. Given the analytical intractability of the received power distribution under spherical wavefronts and diffraction, we propose a subarray partitioning algorithm that accounts for the gradual power change from spherical waves \cite{9940939} while treating diffraction minima as transient noise.

\subsubsection{Power-adaptive subarray segmentation (PASS) algorithm}
We use a sliding-window detection method to capture local power statistics across the array. This involves a window of $W$ elements, within which the power distribution (barring abrupt changes) is assumed to be roughly symmetric. Consequently, statistics measures within the window effectively reflect the local array's power distribution. The window slides across the array, and the power of its trailing element is tested for outliers at each step.
\par To determine the outlyingness of the last window element's power relative to the window statistics, the Mahalanobis distance is introduced:
\begin{equation}
\label{ma_dis}
D_{\text {mal}}^{\text {last}} = \sqrt{\left(p_{\text {last }}-\mu_w\right)^{\mathrm{T}} \Sigma_w^{-1}\left(p_{\text {last }}-\mu_w\right)},
\end{equation}
where $p_{\text{last}}$ is the power of the window's last element, $\mu_w$ and $\Sigma_w$ are the window's mean and variance, respectively. The Mahalanobis distance $D_{\text {mal }}^{\text {last }}$ between $p_{\text{last}}$ and the power distribution of this window is then used as the score distance (SD) for $p_{\text{last}}$.
\par We utilize the reweighted minimum covariance determinant estimator (R-MCD) to estimate the mean and variance of the power distribution within the window. First, a subset of size $h$ is selected within the window, which satisfies $(W+2) / 2 \leq h \leq W$. Among all subsets of size $h$, the subset corresponding to the minimum determinant of the covariance matrix $\operatorname{det}\left(\Sigma_h\right)$ is chosen and denoted as $\mathcal{H}_0$. The mean $\hat{\mu}_0$ and covariance matrix $\hat{\Sigma}_0$ of $\mathcal{H}_0$ are then computed:
\begin{equation}
\label{h_m&v}
\hat{\mu}_0 = \frac{1}{h} \sum_{i \in \mathcal{H}_0} p_i, \quad \hat{\Sigma}_0 = \frac{c_0}{h-1} \sum_{i \in \mathcal{H}_0}\left(p_i-\hat{\mu}_0\right)^2,
\end{equation}
where $c_0=\frac{h / W}{\mathrm{P}\left[\chi^2(3)<\chi_{h / n}^2(1)\right]}$ is a consistency factor.

\par Next, the initial estimates are reweighted as follows:
\begin{equation}
\label{MCD_m&s}
\begin{gathered}\hat{\mu}_{M C D}=\frac{\sum_{i=1}^W p_i I\left[d_i^2<\chi_{0.975}^2(1)\right]}{\sum_{i=1}^W I\left[d_i^2<\chi_{0.975}^2(1)\right]}, \\ \hat{\Sigma}_{M C D}=\frac{c_1}{n-1} \sum_{i=1}^W I\left[d_i^2<\chi_{0.975}^2(1)\right] \cdot\left(p_i-\hat{\mu}_{M C D}\right)^2,
\end{gathered}
\end{equation}
where $d_i=\left(p_i-\hat{\mu}_0\right) / \sqrt{\hat{\Sigma}_0}$, $c_1$ is a consistency factor, and $I(\cdot)$ is an indicator function. $\chi_\alpha^2(i)$ denotes the lower $\alpha$-quantile of the chi-square distribution with $i$ degrees of freedom (DoFs). 
\par In summary, the SD is rewritten as:
\begin{equation}
\label{rew_Dmal}
D_{\text {mal}}^{\text {last}} = \sqrt{\left(p_{\text {last }}-\hat{\mu}_{M C D}\right)^{\mathrm{T}} {\hat{\Sigma}_{M C D}}^{-1}\left(p_{\text {last }}-\hat{\mu}_{M C D}\right)}.
\end{equation}
\par By incorporating the outlier discrimination threshold from robust principal component analysis (ROBPCA), the truncation threshold for the SD is set as $c_\text{SD}=\sqrt{\chi_{0.975}^2(1)}$. When $D_{\text {mal }}^{\text {last }}>c_\text{SD}$, the last array element in the current window is identified as a potential outlier. This threshold is derived from the fact: If $p_{\text {last }} \sim \mathcal{N}\left(\hat{\mu}_{M C D}, \hat{\Sigma}_{M C D}\right)$, then$\left(D_{\text {mal }}^{\text {last }}\right)^2 \sim \chi^2(1)$. Therefore, the rejection region $\mathcal{W}_0=\left\{p_{\text {last }} \mid D_{\text {mal }}^{\text {last }} \geq \sqrt{\chi_\alpha^2(1)}\right\}$ can be defined to reject the hypothesis $p_{\text {last }} \sim \mathcal{N}\left(\hat{\mu}_{M C D}, \hat{\Sigma}_{M C D}\right)$.
\par The sliding window traverses the array, logging detected anomaly indices to form the sequence $\left\{d_n\right\}_{n=1}^N$: 
\begin{equation}
\label{d_n}
d_n= \begin{cases}1, & n \in\left\{n: D_{\text {mal }}^{\text {last }}(n)>c_\text{SD}, n=1,2, \ldots, N\right\} \\ 0, & \text { others }\end{cases}.
\end{equation}
\par To mitigate false positives (FP) and false negatives (FN), based on the array size. we develop a $\frac{W}{4} / \frac{W}{2}$ threshold rule for anomaly detection. Specifically, when a suspicious element is detected, the algorithm calculates the outlier sum (os) over the subsequent $W/2$ elements by summing $d_n$: 
\begin{equation}
\label{os}
    \text{os}_n =\sum_{i=n}^{n+\frac{W}{2}-1}d_n, n=1,\ldots,N
\end{equation}
\par An element is considered a reliable anomaly if $\text{os}_n$ is at least $W/4$:
\begin{equation}
\label{ra}
    n_\text{ra}\in\{i|d_i=1\}\cap\{j|\text{os}_j\geq W/4\},\ i,j=1,\ldots,N.
\end{equation}
\par This rule is grounded in the properties of the R-MCD, which inherently estimates using only the smallest 75\% of anomaly values within the window. $W/4$ represents the number of new points entering the current window that would significantly impact the mean and variance. Thus, if fewer than $W/4$ elements from the next subarray enter, the window's last element is typically flagged as a suspicious anomaly. This validation helps filter transient noise (e.g., diffraction minima and hardware noise) and reliably detect persistent anomalies.
\par
\begin{figure}[!t]
\centering
\includegraphics[width=3.3in]{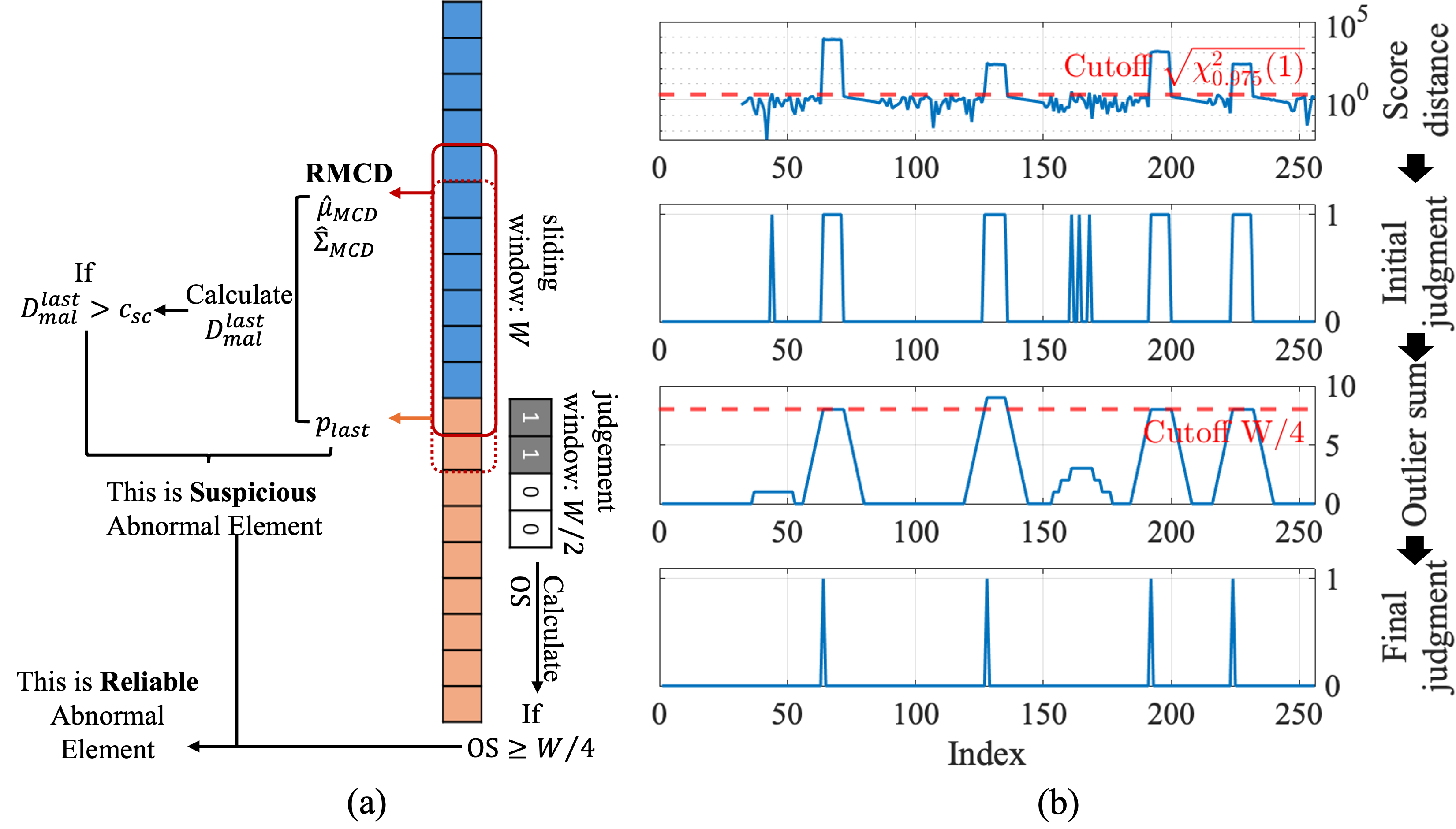}
\caption{PASS algorithm (a) Diagram of PASS, (b) Illustrative example of PASS steps.}
\label{fig_4}
\end{figure}

 The detected SnS birth-death points are sorted as:
\begin{equation}
\label{sort}
\hat{\mathcal{C}} = \operatorname{sort}(\mathcal{C})=\left\{c_0, c_1, c_2, \ldots, c_{N_s-1}, c_{N_s}\right\},
\end{equation}
where $1=c_0<c_1<\cdots<c_{N_S}=N+1$. Here, $N_S$ denotes the number of subarrays obtained from the segmentation. The index set of array elements contained in the $n_s$-th subarray is given by $\mathcal{P}_{n_s}=\left\{c_{n_s-1}, c_{n_s-1}+1, \ldots, c_{n_s}-1\right\}$. The specific implementation of the above process is illustrated in Fig. \ref{fig_4}(a). Furthermore, the complete workflow of PASS algorithm is presented as Fig. \ref{fig_4}(b), which is summarized in Algorithm \ref{alg1}.

\par The PASS algorithm's complexity stems from the MCD and the SD computation. MCD's complexity exhibits sublinear scaling in data size. During the sliding window process, the window performs MCD and SD calculations $N$ times, respectively, leading to a complexity of $\mathcal{O}\left(N(1+\log W)\right)$.

\begin{algorithm}[!b]
\caption{PASS algorithm}\label{alg1}
\renewcommand{\algorithmicrequire}{\textbf{Input:}}
\renewcommand{\algorithmicensure}{\textbf{Output:}}
\begin{algorithmic}[1]
\REQUIRE {Measured power of each array element $\left\{p_n\right\}_{n=1}^N$}
\ENSURE {Detected SnS birth-death points set $\hat{\mathcal{C}}$}
\STATE {\textbf{Initialization}: Place a sliding window of size $W$ at the power sequence start; Set $\hat{\mathcal{C}}$ to the empty set;}
\WHILE{window slides to sequence end}
\STATE {Construct $\mathcal{H}_0$ and compute $\hat{\mu}_0$ and $\hat{\Sigma}_0$ by (\ref{h_m&v});}
\STATE {Reweight $\hat{\mu}_0$ and $\hat{\Sigma}_0$ to obtain $\hat{\mu}_{MCD}$ and $\hat{\Sigma}_{MCD}$ by (\ref{MCD_m&s});}
\STATE {Calculate the Mahalanobis distance of the window's last element to the window by (\ref{rew_Dmal});}
\STATE {Store the obtained value sequentially into $\mathcal{D}_{m}$;}
\ENDWHILE
\STATE {\textbf{Initial judgment}: Convert $\mathcal{D}_{m}$ into the indicator sequence $\left\{d_n\right\}_{n=1}^N$ using the threshold from (\ref{d_n});}
\STATE {Calculate the outlier sum with a window size of $W/2$ and store it in $\left\{\text{os}_n\right\}_{n=1}^N$ by (\ref{os});}
\STATE {\textbf{Final judgment}: Apply $\frac{W}{4}/\frac{W}{2}$ rule to discriminate $\left\{\text{os}_n\right\}_{n=1}^N$ by (\ref{ra}), sort the finally identified birth-death points, and store them into $\hat{\mathcal{C}}$.}
\end{algorithmic}
\end{algorithm}

\subsubsection{Performance}
\par {To quantitatively evaluate subarray segmentation accuracy, we convert Eq. (\ref{sort}) into a binary indicator sequence and compare it against the ground-truth SnS birth–death points. The area under the curve (AUC) derived from this comparison is used as the performance metric, with values in the range $[0.5, 1]$ \cite{BRADLEY19971145}.} Elevated AUC scores demonstrate enhanced discriminative power in identifying true versus false SnS birth-death points, directly quantifying detection fidelity.
\par {The simulation setup is configured as follows: the antenna number is set to $N=256, 512$, the minimum SI is $\text{SI}_\text{min}=32,64,128$, and $L=3$ paths for each UE. We evaluate performance across different SNRs, numbers of UEs $K$, the diffraction intensity factor $t_d$, and the sliding-window size of PASS. For comparison, we benchmark against the rising-and-falling-edges method (RFEM) and the accumulation-function method (AFM) from \cite{9777939}. Specifically, RFEM detects SnS birth–death points by finding peaks in the first-order difference of the power sequence, whereas AFM locates them by identifying slope changes in the accumulation function.}

\par
\begin{figure*}[!t]
\centering
\includegraphics[width=7.25in]{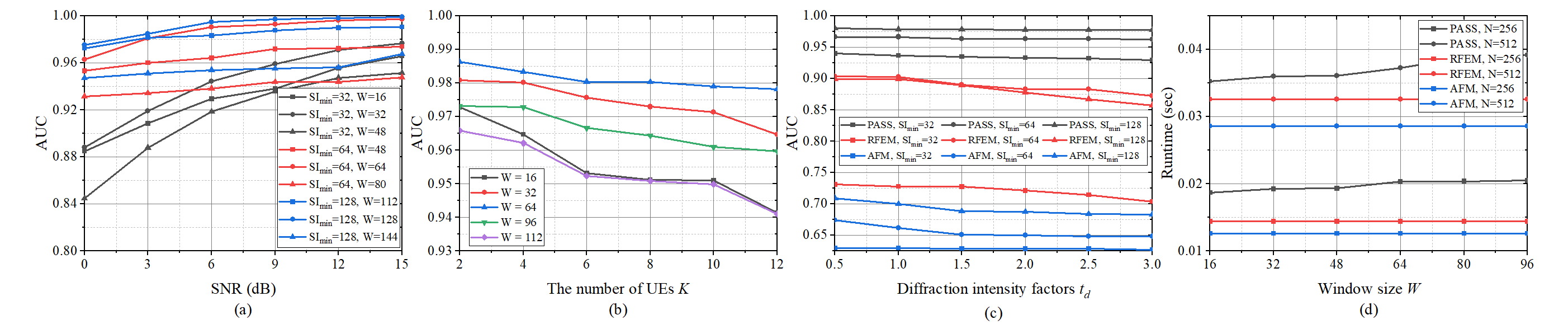}
\caption{{Performance of subarray segmentation. (a) AUC versus SNR: $t_d=1.5$, $K=6$, (b) AUC versus $K$: SNR = 5dB, $t_d=1.5$, (c) AUC versus $t_d$: SNR = 5dB, $K=6$, (d) Runtime of algorithms.}}
\label{auc2}
\end{figure*}

 {Fig. \ref{auc2} presents 500 Monte Carlo simulations. Fig. \ref{auc2}(a) shows that higher SNR improves PASS performance. Increasing $\text{SI}_\text{min}$ likewise stabilizes the spatial-domain channel and yields further gains. Fig. \ref{auc2}(b) shows a slight performance decline as the number of UEs increases, since additional UEs introduce non-ideal paths and amplify power fluctuations. Both figures also reveal small variations in AUC for different sliding window sizes $W$: performance improves as $W$ approaches $\text{SI}_\text{min}$. A too-small $W$ yields insufficient power samples and hampers the distinction between true edge points and power fluctuations. Whereas an overly large $W$ may contain multiple edge points and alter the window statistics. Based on our experiments, we recommend setting $W$ in the range $32$–$64$.}
\par {Fig. \ref{auc2}(c) illustrates that increasing $t_d$ leads to stronger power fluctuations, but PASS algorithm remains robust and outperforms the baseline methods in accuracy. Fig. \ref{auc2}(d) indicates that this accuracy gain incurs only a marginal increase in runtime. Overall, PASS delivers accurate and robust performance across a wide range of channel conditions and parameter settings.}

\subsection{DHBF Architecture-Assisted Array Partitioning \& RF Chain Resource Allocation}
\label{DHBF_sec}
{After obtaining the subarray extents via the PASS algorithm, the subarrays must be decoupled before channel estimation. We introduce the DHBF framework to facilitate subarray decoupling, and its advantages for subarray-based signal processing under SnS are briefly discussed in \cite{9170651}.} When employing the DHBF architecture, subarrays with received power below $\eta_l$ are set to off-mode to optimize resources \cite{10845800}. Let $\hat{N}_S$ denote the number of on-mode subarrays. For simplicity, we will not distinguish between $\hat{N}_S$ and $N_S$ in the following discussion.
\par Consider an XL-MIMO communication system based on uplink time-division duplex (TDD), where $M$ subcarriers serve $K$ single-antenna UEs. The BS is equipped with a DHBF architecture comprising $N_\text{RF}$ RF chains and a ULA with $N$ antenna elements. Due to orthogonal pilot sequences in the CE phase {\cite{tse2005fundamentals}}, we focus on an arbitrary UE. The received signal at the BS for the $m$-th subcarrier at the $p$-th time slot, denoted as $\mathbf{y}_{m, p} \in \mathbb{C}^{N_\text{RF} \times 1}$, can be expressed as
\begin{equation}
\label{system1}
\mathbf{y}_{m, p}=\mathbf{U}_p \mathbf{h}_m v_{m, p}+\mathbf{U}_p \mathbf{n}_{m, p} ,
\end{equation}
where $\mathbf{h}_m \in \mathbb{C}^{N \times 1}$ denotes the channel matrix for the $m$-th subcarrier, $\mathbf{U}_p \in \mathbb{C}^{N_\text{RF} \times N}$ represents the combiner matrix at the BS for the $p$-th time slot, $v_{m, p}$ is the pilot symbol transmitted by the UE at the $p$-th time slot. Without loss of generality, it is assumed that $v_{m, p}=1,\forall m,p$. And $\mathbf{n}_{m, p} \in \mathbb{C}^{N \times 1}$ is the additive white Gaussian noise.
\par For the DHBF architecture, we reformulate the combiner matrix as: 
\begin{equation}
\label{system2}
\mathbf{U}_p=\left[\mathbf{U}_{p, 1}^{\mathrm{T}}, \mathbf{U}_{p, 2}^{\mathrm{T}}, \ldots, \mathbf{U}_{p, N_\text{RF}}^{\mathrm{T}}\right]^{\mathrm{T}},
\end{equation}
where $\mathbf{U}_{p, n_{\mathrm{RF}}} \in \mathbb{C}^{1 \times N}$ represents the combiner matrix corresponding to the $n_\text{RF}$-th RF chain under a fully-connected architecture. Similarly, $\mathbf{H}_m=\left[\mathbf{h}_{m, 1}^{\mathrm{T}}, \ldots, \mathbf{h}_{m, N_{\mathrm{RF}}}^{\mathrm{T}}\right]^{\mathrm{T}}$ and $\mathbf{n}_{m, p}=\left[\mathbf{n}_{m, p, 1}^{\mathrm{T}}, \ldots, \mathbf{n}_{m, p, N_{\mathrm{RF}}}^{\mathrm{T}}\right]^{\mathrm{T}}$. 
\par Since each RF chain is connected to a subset of antenna elements with subarrays as the basic unit, let $\mathcal{S}^{n_{\mathrm{RF}}}$ denote the subarray indices connected to the $n_\text{RF}$-th RF chain (cardinality $N_S^{n_{\mathrm{RF}}}=\left|\mathcal{S}^{n_{\mathrm{RF}}}\right|$) with corresponding antenna indices set $\mathcal{J}_{n_\text{RF}}$ ($N_{n_\text{RF}} = \left| \mathcal{J}_{n_\text{RF}} \right|$). Therefore, the combiner matrix corresponding to the $n_\text{RF}$-th RF chain should be expressed as ${\mathbf{U}}_{p, n_{\mathrm{RF}}}=\mathbf{U}_{p}\left(n_{\mathrm{RF}},\mathcal{J}_{n_{\mathrm{RF}}}\right)$ \footnote{Follow Matlab syntax: Given index sets $\mathcal{A}$ and $\mathcal{B}$, $\mathbf{v}(\mathcal{A})$ extracts elements of vector $\mathbf{v}$ according to the order in $\mathcal{A}$, and $\mathbf{M}(\mathcal{A}, \mathcal{B})$ forms a new matrix by selecting rows of matrix $\mathbf{M}$ indexed by $\mathcal{A}$ and columns by $\mathcal{B}$.}. Similarly, we have ${\mathbf{h}}_{m, n_{\mathrm{RF}}}=\mathbf{h}_{m}\left(\mathcal{J}_{n_{\mathrm{RF}}}\right)$ and ${\mathbf{n}}_{m, p, n_{\mathrm{RF}}}=\mathbf{n}_{m, p}\left(\mathcal{J}_{n_{\mathrm{RF}}}\right)$. 

\par For the $n_\text{RF}$-th RF chain, the received signal at the $p$-th symbol time on the $m$-th subcarrier is given by:
\begin{equation}
\label{rx_y}
y_p^{m, n_{\mathrm{RF}}}={\mathbf{U}}_{p, n_{\mathrm{RF}}} {\mathbf{h}}_{m, n_{\mathrm{RF}}}+{\mathbf{U}}_{p, n_{\mathrm{RF}}} {\mathbf{n}}_{m,p, n_{\mathrm{RF}}}.
\end{equation}
\par The DHBF architecture enables autonomous RF chain-to-element mapping. Under practical $N_{\mathrm{RF}}$ constraints, two operational modes emerge:  when $N_S \leq N_{\mathrm{RF}}$, each subarray exclusively maps to at least one dedicated RF chain and ensures no coupling, i.e., $\left\{N_S^{n_{\mathrm{RF}}}\right\}_{n_{\mathrm{RF}}=1}^{N_\text{RF}} \leq 1$.
\par Conversely, when $N_S>N_\text{RF}$, multiple subarrays must share RF chains, i.e., $\left\{N_S^{n_{\mathrm{RF}}}\right\}_{n_{\mathrm{RF}}=1}^{N_\text{RF}} > 1$, which introduces coupling. To mitigate this, we propose a subarray segmentation-based sampling method (SS-SM). Specifically, for the combiner matrix associated with the $n_\text{RF}$-th RF chain, we employ an indicator matrix to selectively sample and construct a joint combiner matrix across multiple time slots: 
\begin{equation}
    \begin{aligned}
&\boldsymbol{\mathfrak{U}}_{(p-1)/N_S^{n_\text{RF}}+1}^{n_\text{RF}}= {\left[{\mathbf{U}}_{p, n_{\mathrm{RF}}}^{\mathrm{T}}, \ldots, {\mathbf{U}}_{p+{N_S^{n_{\mathrm{RF}}}-1, n_{\mathrm{RF}}}}^{\mathrm{T}}\right]^{\mathrm{T}}} \\
=&  \left[\mathbf{1}_{\{S^{n_\text{RF}}\{1\}\}} \odot {\mathbf{U}}_{p,n_{\mathrm{RF}}}^\text{T},  \ldots,  \mathbf{1}_{\{S^{n_\text{RF}}\{N_S^{n_\text{RF}}\}\}}\odot {\mathbf{U}}_{p,n_{\mathrm{RF}}}^\text{T}\right]^\text{T},
\end{aligned}
\end{equation}
where $\mathcal{S}^{n_\text{RF}}\{l\}$ represents the antenna element index set of the $l$-th subarray connected to the $n_\text{RF}$-th RF chain. And $\mathbf{1}_{\{\mathcal{I}\}}$ denotes indicator vector for  index $\mathcal{I}$.
\par Based on the above equation, for the $n_\text{RF}$-th RF chain, the SS-SM requires $N_S^{n_\text{RF}}$ pilot symbols to achieve one complete decoupling of its connected subarray. Given a total of $P$ pilot symbols, each subarray can be allocated $P/N_S^{n_\text{RF}} \in \mathbb{Z}^+$ effective pilots. The observed signal can be expressed as:
\begin{equation}
    \begin{aligned}
&\left[\hat{y}_{1, 1}^{m, n_\text{RF}},\ldots, \hat{y}_{1, N_S^{n_\text{RF}}}^{m, n_\text{RF}},\ldots,\hat{y}_{P/N_S^{n_\text{RF}}, 1}^{m, n_\text{RF}},\ldots, \hat{y}_{P/N_S^{n_\text{RF}}, N_S^{n_\text{RF}}}^{m, n_\text{RF}}\right]^\text{T}
\\ =& \left[\mathbf{h}_{m,n_\text{RF}}^\text{T}\boldsymbol{\mathfrak{U}}_{1,n_\text{RF}}^\text{T},\ldots,\mathbf{h}_{m,n_\text{RF}}^\text{T}\boldsymbol{\mathfrak{U}}_{P/N_S^{n_\text{RF}},n_\text{RF}}^\text{T}\right]^\text{T}.
\end{aligned}
\end{equation}
\par And the decoupled received signal of the $l$-th subarray for the $n_\text{RF}$-th RF chain is 
\begin{equation}
    \hat{\mathbf{y}}_l^{m, n_\text{RF}}=\left[\hat{y}_{1, l}^{m, n_\text{RF}},  \ldots, \hat{y}_{P / N_S^{n_\text{RF}}, l}^{m, n_\text{RF}}\right]^{\mathrm{T}}, l = 1,\ldots,N_S^{n_\text{RF}}.
\end{equation}
\par Due to SS-SM's property, increasing $N_S^{n_\text{RF}}$ for an RF chain diminishes the effective received signal, resulting in varying effective pilots across subarrays connected to different RF chains. {Since the effect of RF chain resources and effective pilot on subarray-level estimation is not explicit \cite{8723310}, optimizing RF chain allocation for overall channel estimation performance is a complex combinatorial problem. Therefore, exploring better dynamic allocation methods is reserved for future work.} We propose a low-complexity max-element-first greedy allocation algorithm (MEF-GAA) for RF chain allocation, which aims to prioritize more effective pilots for larger subarrays:
\begin{enumerate}
\item{\textbf{Initialization}: Let $\mathcal{C}=\left\{\mathcal{C}_1, \ldots, \mathcal{C}_{N_\text{RF}}\right\}$ be $N_\text{RF}$ empty classes, where each class $\mathcal{C}_i$ is initially an empty set. Let $\left\{\mathcal{P}_1, \ldots, \mathcal{P}_{N_S}\right\}$ be $N_S$ subarrays, where each subarray $\mathcal{P}_j$ contains $n_j$ elements, and $n_1 \geq \cdots \geq n_{N_S}$.}
\item{\textbf{Allocation}: For each subarray $\mathcal{P}_j$ (sorted in descending order of $n_j$): Assign $\mathcal{P}_j$ to the class $\mathcal{C}_k$ that currently has the fewest elements, i.e., $\mathcal{C}_k=\mathcal{C}_k \cup \mathcal{P}_j$, and update the number of elements in class $\mathcal{C}_k$: $\left|\mathcal{C}_k\right|=\left|\mathcal{C}_k\right|+n_j$. Repeat the above steps until all subarrays $\mathcal{P}_j$ have been assigned, and each RF chain $\mathcal{C}_i$ is connected to at least one valid subarray.}
\end{enumerate}
\par {Within the DHBF-PSSP framework, we first determine subarray extents by PASS algorithm. Next, we allocate RF-chain resources based on the number of subarrays and RF chains. For subarrays sharing the same RF chain, we apply SS-SM to decouple received signals. And finally, we will perform subarray channel estimation using the decoupled effective observations, as detailed in Section \ref{sec4}.}

\section{A Near-Field Channel Estimation Method Based on the DFT Codebook} \label{sec4}
Polar-domain codebooks require significant storage and computation. Moreover, varying apertures of partitioned subarrays result in different Rayleigh distances. Thus, uniformly applying polar-domain codebooks across all subarrays would inevitably waste resources and reduce efficiency. Studies in \cite{10638078,10541333} have shown that although the angular-domain sparsity of near-field channels is somewhat diminished, the energy leakage remains confined to a limited range. This characteristic ensures that the angular-domain channel exhibits weak sparsity in XL-MIMO systems, which can be regarded as a block-sparse pattern \cite{arxiv1}.
\par Based on the preceding analysis, this paper proposes a CE algorithm using DFT codebooks to exploit both angular-domain block sparsity and inter-subcarrier structured sparsity in near-field channels. The developed SS-ABSBL-MMV algorithm uniquely integrates structured sparsity constraints to improve estimation accuracy and computational efficiency by leveraging near-field channel properties.
\subsection{SS-ABSBL-MMV Assisted by Block-Structured Sparsity}
\begin{figure}[!t]
\centering
\includegraphics[width=3.4in]{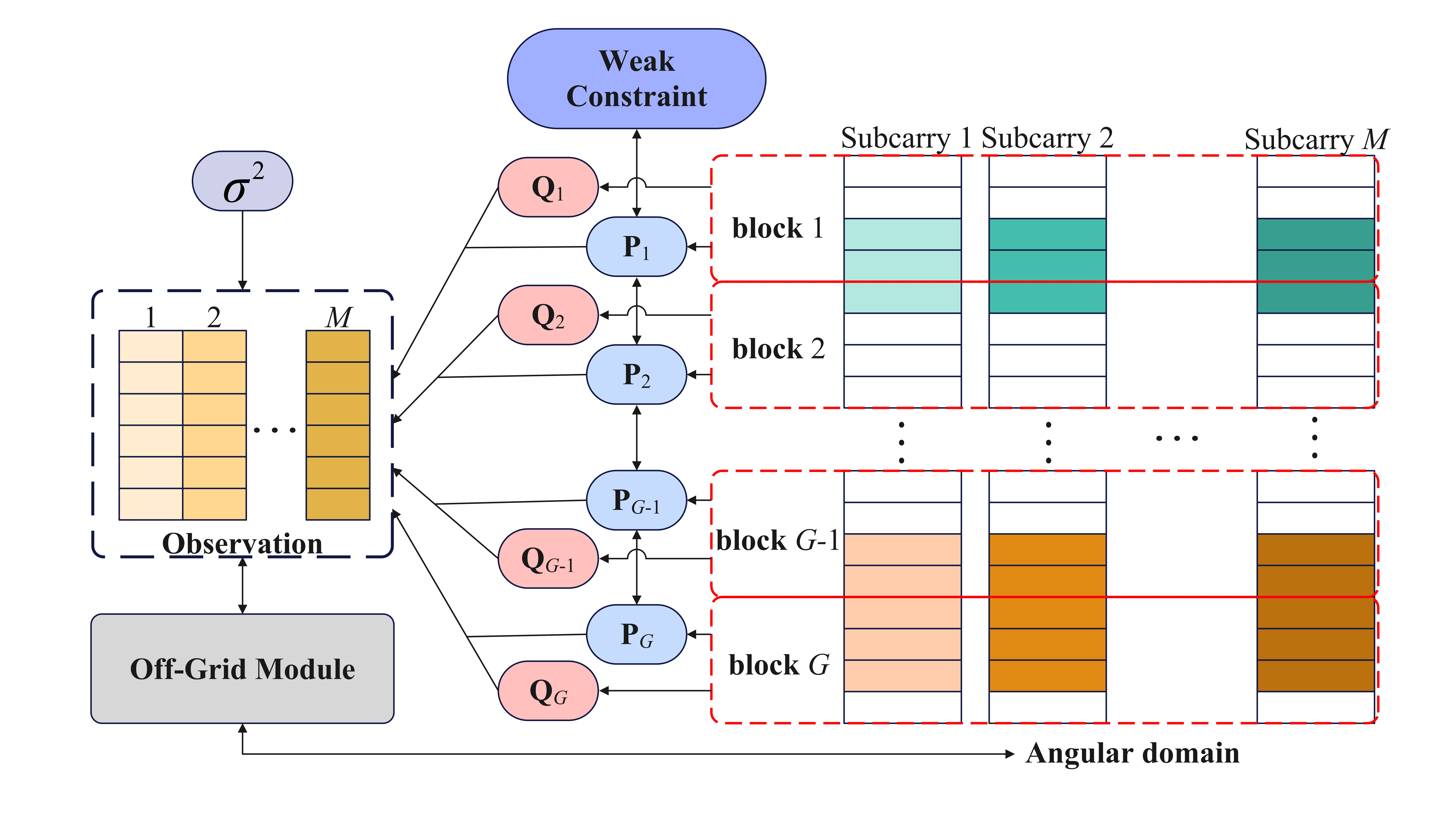}
\caption{Diagram of ABSBL using MMV framework and Off-grid module.}
\label{fig_7}
\end{figure}
\par For $n_s$-th subarray after decoupling, here is a rephrased version of the effective received signal at the $p_{n_s}$-th pilot time for the $m$-th subcarrier:
\begin{equation}
\label{absbl1}
\mathbf{y}_{m, p}^{n_s}=\boldsymbol{\Phi}_p^{n_s} \mathbf{h}_m^{n_s} +\mathbf{n}_{m, p}^{n_s},
\end{equation}
where $\boldsymbol{\Phi}_p^{n_s}\in \mathbb{C}^{N_\text{RF}^{n_s} \times N_{n_s}}$ denotes the combiner matrix for $n_s$-th subarray at the $p_{n_s}$-th pilot time, $\mathbf{h}_m^{n_s}$ represents the channel matrix under the $m$-th subcarrier. And $\mathbf{n}_{m, p}^{n_s} \in \mathbb{C}^{N_\text{RF}^{n_s} \times 1}$ is the additive white Gaussian noise following $\mathcal{C N}\left(\mathbf{0}, \sigma_n^2 \mathbf{I}\right)$.\footnote{{To keep the notation concise, this chapter drops the subscript $n_s$ and treats subarrays generically.}} Therefore, the received signals at the $m$-th subcarrier for all $P$ pilot times can be written as:
\begin{equation}
\label{absbl2}
\mathbf{y}_m=\boldsymbol{\Phi} \mathbf{h}_m+\mathbf{n}_m,
\end{equation}
where $\mathbf{y}_m=\left[\mathbf{y}_{m, 1}^{\mathrm{T}}, \ldots \mathbf{y}_{m, P}^{\mathrm{T}}\right]^{\mathrm{T}} \in \mathbb{C}^{P N_{\mathrm{RF}} \times 1}$, $\boldsymbol{\Phi}=\left[\boldsymbol{\Phi}_1^{\mathrm{T}}, \ldots, \boldsymbol{\Phi}_P^{\mathrm{T}}\right]^{\mathrm{T}} \in \mathbb{C}^{P N_{\mathrm{RF}} \times N}$, and $\mathbf{n}_m=\left[\mathbf{n}_{m, 1}^{\mathrm{T}}, \ldots \mathbf{n}_{m, P}^{\mathrm{T}}\right]^{\mathrm{T}} \in \mathbb{C}^{P N_{\mathrm{RF}} \times 1}$. Then, the received signals across all $M$ subcarriers can be expressed in a more compact form as:
\begin{equation}
\label{absbl3}
\mathbf{Y}=\boldsymbol{\Phi} \mathbf{H}+\mathbf{N}.
\end{equation}

Among them, $\mathbf{Y}=\left[\mathbf{y}_1, \ldots, \mathbf{y}_M\right] \in \mathbb{C}^{P N_{\mathrm{RF}} \times M}$, $\mathbf{H}=\left[\mathbf{h}_1, \ldots, \mathbf{h}_M\right] \in \mathbb{C}^{N \times M}$, and $\mathbf{N}=\left[\mathbf{n}_1, \ldots, \mathbf{n}_M\right] \in \mathbb{C}^{P N_{\mathrm{RF}} \times M}$. The channel matrix is represented in the angular domain as:
\begin{equation}
\label{absbl4}
\mathbf{Y}=\boldsymbol{\Psi} \mathbf{X}+\mathbf{N},
\end{equation}
where $\mathbf{X}=\left[\mathbf{x}_1, \ldots, \mathbf{x}_M\right] \in \mathbb{C}^{N \times M}$ is the angular-domain channel matrix. Let the sensing matrix $\boldsymbol{\Psi}=\boldsymbol{\Phi} \mathbf{D}$.  Here $\mathbf{D}$ is the DFT dictionary matrix. 
\par Assuming that $\mathbf{x}_m$ across subcarriers are independent and follow $\mathcal{C N}\left(\mathbf{0}, \mathbf{V}_m\right)$. To accurately capture the unique block sparsity characteristic of the near-field channel in the angular domain, we partition $\mathbf{x}_m$. {Without loss of generality, we assume that $\mathbf{x}_m$ is divided into $G$ blocks of the same length $U$, and $N=GU$.} Therefore, $\mathbf{x}_m$ can be written as: 
\begin{equation}
\label{dbabl6}
\mathbf{x}_m=[\underbrace{x_{1,1}^m, x_{1,2}^m, \ldots, x_{1, U}^m}_{1\text {-st block } \mathbf{x}_1^m}, \ldots, \underbrace{x_{G, 1}^m,\ldots, x_{G, U}^m}_{G\text {-th block } \mathbf{x}_G^m}]^{\mathrm{T}}.
\end{equation}

\par To capture the intra-block diversity and inter-block correlation, we model the prior distribution of the $g$-th block $\mathbf{x}_g^m$ as \cite{NEURIPS2024_ead542f1}:
\begin{equation}
\label{absbl7}
p\left(\mathbf{x}_g^m ;\left\{\mathbf{Q}_g^m, \mathbf{P}_g^m\right\}\right)=\mathcal{C N}\left(\mathbf{0}, \mathbf{Q}_g^m \mathbf{P}_g^m \mathbf{Q}_g^m\right),
\end{equation}
where $\mathbf{Q}_g^m=\operatorname{diag}\left\{\sqrt{\gamma_{g, 1}^m}, \ldots, \sqrt{\gamma_{g, K}^m}\right\}$ is used to capture the intra-block variance diversity, and $\mathbf{P}_g^m$ is used to capture the inter-block variance correlation, which is composed of $\iota_{i j}^m$ for all $i,j=1,…,U$. By assembling all the blocks, we can obtain $\mathbf{V}_m=\operatorname{diag}\left\{\mathbf{Q}_1^m \mathbf{P}_1^m \mathbf{Q}_1^m, \ldots, \mathbf{Q}_G^m \mathbf{P}_G^m \mathbf{Q}_G^m\right\}$. Notably, if $\mathbf{Q}_g^m=\gamma_g^m\mathbf{I}_U,\forall g,m$, the model reduces to the conventional BSBL framework in \cite{8590774}.
\par In multi-carrier systems, the near-field angular-domain channel sparsity exhibits a structured characteristic, i.e., the same block support structure exists under different subcarriers $f_m$, a feature referred to as BMMV \cite{arxiv1}. For each block $\mathbf{x}_g^m$, they share the same covariance matrix across different subcarriers $f_m$, i.e., $\mathbf{Q}_g^m \mathbf{P}_g^m \mathbf{Q}_g^m=\mathbf{Q}_g \mathbf{P}_g \mathbf{Q}_g$ ($\forall m$).
\par To elucidate the BMMV problem, the original problem is reformulated as:
\begin{equation}
\label{absbl8}
\tilde{\mathbf{y}}=\widetilde{\mathbf{\Psi}} \tilde{\mathbf{x}}+\widetilde{\mathbf{n}}.
\end{equation}

Among them, $\tilde{\mathbf{y}}=\operatorname{vec}\left(\mathbf{Y}^{\mathrm{T}}\right) \in \mathbb{C}^{P M N_{\mathrm{RF}} \times 1}$, $\widetilde{\boldsymbol{\Psi}}=\boldsymbol{\Psi} \otimes \mathbf{I}_M \in \mathbb{C}^{P M N_{\mathrm{RF}} \times M N}$, $\tilde{\mathbf{x}}=\operatorname{vec}\left(\mathbf{X}^{\mathrm{T}}\right) \in \mathbb{C}^{M N \times 1}$, and $\widetilde{\mathbf{n}}=\operatorname{vec}\left(\mathbf{N}^{\mathrm{T}}\right) \in \mathbb{C}^{P M N_{\mathrm{RF}} \times 1}$. Then, for the reconstructed $\widetilde{\mathbf{x}}$, it can still be divided into $G$ blocks, with the channel vector in each block being $\tilde{\mathbf{x}}_g=\left[x_{g, 1}^1, \ldots, x_{g, U}^1, \ldots, x_{g, 1}^M, \ldots, x_{g, U}^M\right]^{\mathrm{T}} \in \mathbb{C}^{M U \times 1}$, then we have:
\begin{equation}
\label{absbl9}
p\left(\tilde{\mathbf{x}}_g ;\left\{\mathbf{Q}_g, \mathbf{P}_g\right\}\right)=\mathcal{C N}\left(\mathbf{0}, \mathbf{I}_M \otimes \mathbf{Q}_g \mathbf{P}_g \mathbf{Q}_g\right).
\end{equation}
\par By assembling all the blocks, the prior distribution of $\widetilde{\mathbf{x}}$ is following $\mathcal{C N}(\mathbf{0}, \widetilde{\mathbf{V}})$, where $\widetilde{\mathbf{V}}=\operatorname{diag}\left\{\mathbf{I}_M \otimes \mathbf{Q}_1 \mathbf{P}_1 \mathbf{Q}_1, \ldots, \mathbf{I}_M \otimes \mathbf{Q}_G \mathbf{P}_G \mathbf{Q}_G\right\}$.
\par Based on the observation of $\widetilde{\mathbf{y}}$, the Gaussian likelihood function can be obtained as:
\begin{equation}
\label{absbl10}
p\left(\tilde{\mathbf{y}} \mid \tilde{\mathbf{x}}, \sigma_n^2\right)=\mathcal{C N}\left(\widetilde{\mathbf{\Psi}} \tilde{\mathbf{x}}, \sigma_n^2 \mathbf{I}\right).
\end{equation}
\par Leveraging the Bayesian estimation principle and the properties of the multivariate Gaussian distribution, the posterior probability distribution can be derived as:
\begin{equation}
\label{absbl11}
p\left(\tilde{\mathbf{x}} \mid \tilde{\mathbf{y}} ;\left\{\mathbf{Q}_g, \mathbf{P}_g\right\}_{g=1}^G, \sigma_n^2\right)=\mathcal{C N}\left(\boldsymbol{\mu}_{\mathbf{x}}, \mathbf{\Sigma}_{\mathbf{x}}\right),
\end{equation}
where
\begin{equation}
\label{absbl12}
\begin{gathered}
\boldsymbol{\mu}_{\mathbf{x}}=\widetilde{\mathbf{V}} \widetilde{\boldsymbol{\Psi}}^{\mathrm{H}}\left(\sigma_n^2 \mathbf{I}+\widetilde{\boldsymbol{\Psi}} \widetilde{\mathbf{V}} \widetilde{\boldsymbol{\Psi}}^{\mathrm{H}}\right)^{-1} \tilde{\mathbf{y}}, \\
\boldsymbol{\Sigma}_{\mathbf{x}}=\widetilde{\mathbf{V}}-\widetilde{\mathbf{V}} \widetilde{\boldsymbol{\Psi}}^{\mathrm{H}}\left(\sigma_n^2 \mathbf{I}+\widetilde{\boldsymbol{\Psi}} \widetilde{\mathbf{V}}\widetilde{\boldsymbol{\Psi}}^{\mathrm{H}}\right)^{-1} \widetilde{\boldsymbol{\Psi}} \widetilde{\mathbf{V}}.
\end{gathered}
\end{equation}

By estimating the hyperparameters $\boldsymbol{\Theta}=\left\{\left\{\mathbf{Q}_g, \mathbf{P}_g\right\}_{g=1}^G, \sigma_n^2\right\}$, the MAP estimate of $\widetilde{\mathbf{x}}$ can be obtained as $\hat{\boldsymbol{\mu}}=\boldsymbol{\mu}_{\mathbf{x}}$.
\par Next, the Expectation-Maximization (EM) algorithm is employed to estimate the hyperparameters $\mathbf{\Theta}$. The main idea of estimating $\mathbf{\Theta}$ using the EM algorithm is to maximize the likelihood function $p(\tilde{\mathbf{y}} , \widetilde{\mathbf{x}}; \boldsymbol{\Theta})$, which is equivalent to:
\begin{equation}
\label{absbl13}
\min _{\boldsymbol{\Theta}} \mathcal{L}(\boldsymbol{\Theta})=-\ln p(\tilde{\mathbf{y}} , \tilde{\mathbf{x}}; \boldsymbol{\Theta})=\tilde{\mathbf{y}}^{\mathrm{H}} \boldsymbol{\Sigma}_{\mathbf{y}}{ }^{-1} \tilde{\mathbf{y}}+\ln \left|\boldsymbol{\Sigma}_{\mathbf{y}}\right|,
\end{equation}
where $\boldsymbol{\Sigma}_{\mathbf{y}}=\sigma_n^2 \mathbf{I}+\widetilde{\boldsymbol{\Psi}} \widetilde{\mathbf{V}} \widetilde{\mathbf{\Psi}}^{\mathrm{H}}$. In the E-step, $\widetilde{\mathbf{x}}$ is treated as a latent variable, and the following $Q$-function is maximized:

\begin{equation}
\label{absbl14}
\begin{split}
Q(\boldsymbol{\Theta}) &= \mathbb{E}_{\tilde{\mathbf{x}} | \tilde{\mathbf{y}}, \boldsymbol{\Theta}^{(t-1)}} [\ln p(\tilde{\mathbf{y}}, \tilde{\mathbf{x}}; \boldsymbol{\Theta})] \\ &= \mathbb{E}_{\tilde{\mathbf{x}} | \tilde{\mathbf{y}}, \boldsymbol{\Theta}^{(t-1)}} [\ln p(\tilde{\mathbf{y}} | \tilde{\mathbf{x}}, \sigma_n^2)] \\ &+ \mathbb{E}_{\tilde{\mathbf{x}} | \tilde{\mathbf{y}}, \boldsymbol{\Theta}^{(t-1)}} \left[ \ln p\left(\tilde{\mathbf{x}}, \{\mathbf{Q}_g, \mathbf{P}_g\}_{g=1}^G \right) \right].
\end{split}
\end{equation}
\par The $Q$-function is divided into parts that solely contain the hyperparameter $\sigma_n^2$ and those that solely contain the hyperparameters $\left\{\mathbf{Q}_g, \mathbf{P}_g\right\}_{g=1}^G$. \par In the M-step, we obtain estimates of $\boldsymbol{\Theta}$ by maximizing the $Q(\boldsymbol{\Theta})$. First, term $\mathbb{E}_{\tilde{\mathbf{x}} | \tilde{\mathbf{y}}, \boldsymbol{\Theta}^{(t-1)}} \left[\ln p\left(\tilde{\mathbf{x}}, \{\mathbf{Q}_g, \mathbf{P}_g\}_{g=1}^G \right) \right]$ can be expressed as:
\begin{equation}
\label{absbl15}
\begin{split}
&\mathbb{E}_{\tilde{\mathbf{x}} | \tilde{\mathbf{y}}, \boldsymbol{\Theta}^{(t-1)}} \left[\ln p\left(\tilde{\mathbf{x}}, \{\mathbf{Q}_g, \mathbf{P}_g\}_{g=1}^G \right) \right] \\ \propto & {-\frac{1}{2}  \ln |\tilde{\mathbf{V}}| -\frac{1}{2} \left[\text{tr} \left( \tilde{\mathbf{V}}^{-1}  \boldsymbol{\Sigma}_{{\mathbf{x}}} \right) + \boldsymbol{\mu}_{{\mathbf{x}}}^\text{H}\tilde{\mathbf{V}}^{-1} \boldsymbol{\mu}_{{\mathbf{x}}}\right]}.
\end{split}
\end{equation}
\par Estimating $\mathbf{Q}_g$ is equivalent to estimating $\left\{ \sqrt{\gamma_{g,u}} \right\}_{u=1}^{U}$. Therefore, to maximize (\ref{absbl15}), we first take the derivative of its first part:
\begin{equation}
\label{absbl16}
\frac{\partial \left( -\frac{1}{2} \ln |\mathbf{V}| \right)}{\partial \sqrt{\gamma_{g,u}}} = \frac{\partial \left( -\frac{M}{2} \sum_{i=1}^G |\mathbf{Q}_i \mathbf{P}_i \mathbf{Q}_i| \right)}{\partial \sqrt{\gamma_{g,u}}} = \frac{-M}{\sqrt{\gamma_{g,u}}}.
\end{equation}

\par Then take the derivative of the second term: 
\begin{equation}
	\label{absbl17}
	\begin{split}
&\frac{\partial \left( -\frac{1}{2} \left[\text{tr} \left( \tilde{\mathbf{V}}^{-1}  \boldsymbol{\Sigma}_{{\mathbf{x}}} \right) + \boldsymbol{\mu}_{{\mathbf{x}}}^\text{H}\tilde{\mathbf{V}}^{-1} \boldsymbol{\mu}_{{\mathbf{x}}} \right] \right)}{\partial \sqrt{\gamma_{g,u}}} \\=& {-\frac{1}{2} \text{tr} \left[ \mathbf{I}_M \otimes \frac{\partial \left( \mathbf{Q}_g \mathbf{P}_g \mathbf{Q}_g \right)^{-1}}{\partial \sqrt{\gamma_{g,u}}} \mathbf{R}^g \right]} \\
\\ =& {\gamma_{g,u}}^{-\frac{3}{2}} \left( \mathbf{P}_g^{-1} \right)_{u,u} \text{tr}\left[\sum_{m=1}^M \left(\mathbf{R}^g\right)_{m}\mathbf{e}_u \mathbf{e}_u^\text{T}\right] 
\\+& {\gamma_{g,u}}^{-1}  \text{tr}\left[(\mathbf{W}_{g \backslash u}\mathbf{P}_g)^{-1}     \sum_{m=1}^M \left(\mathbf{R}^g\right)_{m}\mathbf{e}_u \mathbf{e}_u^\text{T}\right],
    \end{split}
	\end{equation}
where $\mathbf{R}^g=\left( \boldsymbol{\Sigma}_{\mathbf{x}}^g + \boldsymbol{\mu}_{\mathbf{x}}^g {\boldsymbol{\mu}_{\mathbf{x}}^{g}}^{\text{H}} \right)$, 
$\boldsymbol{\mu}_{\mathbf{x}}^g = \boldsymbol{\mu}_{\mathbf{x}}((g-1)MU+1:gMU)$ and $\boldsymbol{\Sigma}_{\mathbf{x}}^g = ((g-1)MU+1:gMU,\quad (g-1)MU+1:gMU)$, while $(\cdot)_{i,j}$ denotes the element at the $\left(i,j\right)$-th index of the matrix. Additionally, $\mathbf{W}_{g\backslash u} = \text{diag}\left\{ \sqrt{\gamma_{g,1}}, \ldots, \sqrt{\gamma_{g,u-1}}, 0, \sqrt{\gamma_{g,u+1}}, \ldots, \sqrt{\gamma_{g,U}} \right\}$. 

\par {By combining the above derivative expressions and setting $\frac{\partial \mathbb{E}_{\tilde{\mathbf{x}} | \tilde{\mathbf{y}}, \boldsymbol{\Theta}^{(t-1)}} \left[ \ln p\left(\tilde{\mathbf{x}}, \{\mathbf{Q}_g, \mathbf{P}_g\}_{g=1}^G \right) \right]}{\partial \sqrt{\gamma_{g,u}}}=0$, we can obtain update rule for $\gamma_{g,k}$:}
\begin{equation}
\label{absbl18}
\gamma_{g,u} = \left(\sqrt{\frac{\mathbf{B}_{g,u}^2}{4M^2}+\frac{\mathbf{A}_{g,u}}{M}}+\frac{\mathbf{B}_{g,u}}{2M}\right)^2.
\end{equation}

In the above equation,
\begin{gather*}
\mathbf{A}_{g,u} = \left( \mathbf{P}_g^{-1} \right)_{u,u} \text{tr}\left[\sum_{m=1}^M \left(\mathbf{R}^g\right)_{m}\mathbf{e}_u \mathbf{e}_u^\text{T}\right], \\ 
\mathbf{B}_{g,u} =  \text{tr}\left[(\mathbf{W}_{g \backslash u}\mathbf{P}_g)^{-1}     \sum_{m=1}^M \left(\mathbf{R}^g\right)_{m}\mathbf{e}_u \mathbf{e}_u^\text{T}\right].
\end{gather*}
where $\left(\mathbf{R}^g\right)_m=\mathbf{R}^g((m-1)U+1: mU,\quad(m-1)U+1:mU)$. Similarly, by taking the derivative of both parts of (\ref{absbl15}) with respect to $\mathbf{P}_g$, we obtain:

\begin{equation}
\label{absbl19}
\begin{split}
    &{\frac{\partial \mathbb{E}_{\tilde{\mathbf{x}} | \tilde{\mathbf{y}}, \boldsymbol{\Theta}^{(t-1)}} \left[ \ln p\left(\tilde{\mathbf{x}}, \{\mathbf{Q}_g, \mathbf{P}_g\}_{g=1}^G \right) \right]}{\partial \mathbf{P}_g}} \\
	=&\frac{\partial \left( -\frac{1}{2} \ln |\tilde{\mathbf{V}}| -\frac{1}{2} \text{tr} \left[ \tilde{\mathbf{V}}^{-1} \left( \boldsymbol{\Sigma}_{\mathbf{x}} + \boldsymbol{\mu}_{\mathbf{x}} \boldsymbol{\mu}_{\mathbf{x}}^\text{H} \right) \right] \right)}{\partial \mathbf{P}_g} \\ =& -\frac{M}{2} \mathbf{P}_g^{-1} + \frac{1}{2} \sum_{m=1}^M \left[ \mathbf{P}_g^{-1} \mathbf{Q}_g^{-1} \left(\mathbf{R}^g\right)_m \mathbf{Q}_g^{-1} \mathbf{P}_g^{-1} \right].
\end{split}
\end{equation}

Combine the above derivation and set it to zero to get:
\begin{equation}
\label{absbl20}
\mathbf{P}_g = \frac{1}{M} \mathbf{Q}_g^{-1} \sum_{m=1}^M \left(\mathbf{R}^g\right)_m \mathbf{Q}_g^{-1}.
\end{equation}
\par {Similarly, we derive the likelihood function for $\sigma_n^2$:}
\begin{equation}
\begin{aligned}
    &{\mathrm{E}_{\mathbf{x} \mid \mathbf{y}, \boldsymbol{\Theta}^{t-1}}\left[\ln p\left(\tilde{\mathbf{y}} \mid \tilde{\mathbf{x}}, \sigma^{2}\right)\right] \propto-\frac{M P N_{\mathrm{RF}}}{2} \ln \sigma_n^{2}} \\
    &{-\frac{1}{2 \sigma_n^{2}}\left\{\left\|\tilde{\mathbf{y}}-\tilde{\boldsymbol{\Psi}} \boldsymbol{\mu}_{\mathbf{x}}\right\|_{2}^{2} +\sigma_{n,(t-1)}^{2}\left[N M-\operatorname{tr}\left(\boldsymbol{\Sigma}_{\mathbf{x}} \tilde{\mathbf{V}}^{-1}\right)\right]\right\}}
\end{aligned}
\end{equation}
\par {Setting the derivative to zero gives the update rule for $\sigma_n^2$:}
\begin{equation}
\label{absbl22}
\sigma_{n,(t)}^2 = \frac{\left\| \tilde{\mathbf{y}} - \tilde{\boldsymbol{\Psi}} \boldsymbol{\mu}_{\mathbf{x}} \right\|_2^2 + \sigma_{n,(t-1)}^2 \left[ NM - \text{tr} \left( \boldsymbol{\Sigma}_{\mathbf{x}} \tilde{\mathbf{V}}^{-1} \right) \right]}{MPN_\text{RF}}.
\end{equation}
\par To prevent overfitting $\mathbf{P}_g$ and ensure inter-block variance diversity, we correct $\mathbf{P}_g$ with constraints. Specifically, we move from unconstrained maximization of $\mathbb{E}_{\tilde{\mathbf{x}} | \tilde{\mathbf{y}}, \boldsymbol{\Theta}^{(t-1)}} \left[\ln p\left( \tilde{\mathbf{x}}, \{\mathbf{Q}_g, \mathbf{P}_g\}_{g=1}^G \right) \right]$ to a constrained one. This paper adopts the weak constraint function $\ln\left[\text{det}(\cdot)\right]$, and thus the original problem becomes \cite{NEURIPS2024_ead542f1}:
\begin{equation}
\label{absbl23}
\begin{aligned}
\max_{\mathbf{P}_g} \quad &-\frac{1}{2} \ln \left[ \text{det}\left(\tilde{\mathbf{V}}\right) \right] - \frac{1}{2} \text{tr} \left[ \tilde{\mathbf{V}}^{-1} \left( \boldsymbol{\Sigma}_{\mathbf{x}} + \boldsymbol{\mu}_{\mathbf{x}} \boldsymbol{\mu}_{\mathbf{x}}^\text{H} \right) \right] \\
\text{s.t.} \quad &\ln \left[ \text{det} \left(\mathbf{I}_M \otimes \mathbf{P}_g\right)\right] = \ln \left[ \text{det} \left(\bar{\mathbf{P}}\right) \right]
\end{aligned},
\end{equation}
where $\bar{\mathbf{P}} = \frac{1}{G} \sum_{g=1}^G (\mathbf{I}_M \otimes \mathbf{P}_g)$.
\par {To solve the problem in (\ref{absbl23}), the augmented Lagrangian method (ALM) can be employed, which involves constructing and minimizing an augmented Lagrangian function:}
\begin{equation}
\begin{aligned}
\min _{\mathbf{P}_{g}} \mathcal{L}\left(\mathbf{P}_{g}, \lambda, c\right)=&{\frac{1}{2} \ln |\widetilde{\mathbf{V}}|+\frac{1}{2} \operatorname{tr}\left[\widetilde{\mathbf{V}}^{-1}\left(\boldsymbol{\Sigma}_{\mathbf{x}}+\boldsymbol{\mu}_{\mathbf{x}} \boldsymbol{\mu}_{\mathbf{x}}^{\mathrm{H}}\right)\right]}\\
+&{\sum_{g=1}^{G} \lambda_{g}\left(\ln \left|\mathbf{I}_{M} \otimes \mathbf{P}_{g}\right|-\ln |\overline{\mathbf{P}}|\right)}\\
+&{\frac{c}{2}\left\|\ln \left|\mathbf{I}_{M} \otimes \mathbf{P}_{g}\right|-\ln |\overline{\mathbf{P}}|\right\|^{2}},
\end{aligned}
\end{equation}
where $\lambda$ is the Lagrange multiplier, and $c$ is the quadratic penalty factor. Solving $\nabla_{\mathbf{P}_g} \mathcal{L}(\mathbf{P}_g, \lambda, c) = 0$ and $\nabla_{\lambda} \mathcal{L}(\mathbf{P}_g, \lambda, c) = 0$ yields the iterative ALM formulas:

\begin{equation}
\label{absbl25}
\begin{aligned}
\mathbf{P}_g^{(t)} &= \frac{\mathbf{Q}_g^{-1} \sum_{m=1}^M \left(\mathbf{R}^g\right)_m \mathbf{Q}_g^{-1}}{M \left[ 1 + 2 \lambda_g^{(t-1)} + 2c \left( M \ln |\mathbf{P}_g^{(t-1)}| - \ln |\bar{\mathbf{P}}^{(t-1)}| \right) \right]} \\
\lambda_g^{(t)} &= \lambda_g^{(t-1)} + \alpha^{(t-1)} \left( M \ln |\mathbf{P}_g^{(t-1)}| - \ln |\bar{\mathbf{P}}^{(t-1)}| \right)
\end{aligned},
\end{equation}
where $\alpha$ is the iteration step size of the ALM, and $\lambda_g$ represents the multiplier associated with the $g$-th block.

\subsection{Introduction of the Off-Grid Module}
The limited resolution of DFT-based angular-sparsity CE causes angle mismatch, leading to biased estimation and increased angular spread errors in near-field scenarios. To address this issue, we seek a set of angles $\boldsymbol{\hat{\Xi}} = [\hat{\theta}_1, \ldots, \hat{\theta}_N]$ and reconstruct the codebook $\mathbf{D}(\boldsymbol{\hat{\Xi}}) = [\mathbf{a}(\hat{\theta}_1), \ldots, \mathbf{a}(\hat{\theta}_N)]$ for channel matrix estimation. For a single subcarrier, the optimization problem is:
\begin{equation}
\label{absbl26}
\mathbf{z}^*, \mathbf{X}^* = \arg\min_{\mathbf{z}, \mathbf{x}} \|\mathbf{Y} - \boldsymbol{\Phi} \mathbf{D}(\arcsin(\mathbf{z})) \mathbf{X}\|^2,
\end{equation}
where $\mathbf{z} = \sin{\mathbf{\Xi}}$.
\par We employ an alternating framework to address the problem in (\ref{absbl26}). Specifically, in each iteration, the sparse coefficient vector is updated first using the least squares (LS) solution:
\begin{equation}
\label{absbl27}
\mathbf{X}^{(t+1)} = \left[ \left( \boldsymbol{\Phi} \mathbf{D}^{(t)} \right)^\text{H} \boldsymbol{\Phi} \mathbf{D}^{(t)} \right]^{-1} \left( \boldsymbol{\Phi} \mathbf{D}^{(t)} \right)^\text{H} \mathbf{Y}.
\end{equation}
\par For the angular grid, the gradient descent method (GD) is employed for updates, with the following computed first:
\begin{equation}
\label{absbl28}
\begin{aligned}
\nabla_{\mathbf{z}} f^{(t+1)}(\mathbf{z}, \mathbf{X}) = 2\Re\left\{j\frac{2\pi d} {\lambda} \text{diag}\left[\mathbf{X}^{(t+1)}\right.\right.\\
\left.\left.\left( \mathbf{Y} - \boldsymbol{\Phi} \mathbf{D}^{(t)} \mathbf{X}^{(t+1)} \right)^\text{H}\boldsymbol{\Phi}\left(\mathbf{L}_\text{ant}\odot\mathbf{D}^{(t)}\right)\right]\right\},
\end{aligned}
\end{equation}
where each column of $\mathbf{L}_\text{ant}$ is $\left[0,\ldots,N-1\right]^\text{T}$.
\par Simultaneously, the update step size of the GD is determined by the backtracking line search approach, and the iterative formula is obtained: 
\begin{equation}
\label{absbl29}
\mathbf{z}^{(t+1)} = \mathbf{z}^{(t)} - \rho^{(t)} \nabla_{\mathbf{z}} f^{(t+1)}(\mathbf{z}, \mathbf{X}),
\end{equation}
where $\rho$ represents the update step size of GD. The codebook is updated via $\mathbf{D}^{(t+1)} = \mathbf{D}(\arcsin(\mathbf{z}^{(t+1)}))$. The finally estimated channel matrix is generated by $\mathbf{D}^{(T)}\mathbf{X}^{(T)}$. 
\par {To reduce complexity and ensure convergence, we threshold the on-grid estimate $\mathbf{X}$ from ABSBL-MMV, and keep only rows with significant entries and their corresponding angular grid points for off-grid processing.} We refer to the SS-ABSBL-MMV algorithm with an off-grid module as SS-OG-ABSBL-MMV. The schematic diagram of this algorithm is shown in Fig. \ref{fig_7}, and the pseudocode is given by Algorithm \ref{alg2}.

\begin{algorithm}[!b]
\caption{ABSBL Algorithm.}\label{alg2}
\renewcommand{\algorithmicrequire}{\textbf{Input:}}
\renewcommand{\algorithmicensure}{\textbf{Output:}}
\begin{algorithmic}[1]
\REQUIRE {Received signal $\mathbf{Y}\in \mathbb{C}^{PN_\text{RF} \times M}$; Sensing matrix $\mathbf{\Psi}=\mathbf{\Phi D} \in \mathbb{C}^{PN_\text{RF} \times N}$; Block size $U=N/G$; ABSBL and off-grid module maximum iteration number $T_{ite}$, $R_{ite}$, and their corresponding stopping thresholds $\delta_1$, $\delta_2$.}
\ENSURE {Estimated channel matrix $\hat{\mathbf{H}}$.}
\STATE {\textbf{Initialization}: Rearrange $\mathbf{Y}$ and $\mathbf{\Psi}$ to $\tilde{\mathbf{y}}$ and $\tilde{\mathbf{\Psi}}$ by (\ref{absbl8}); Set $\mathbf{Q}_g, \mathbf{P}_g = \mathbf{I}_U, \forall g$; Initial noise variance $\sigma_{(0)}^2 = \text{var}(\tilde{\mathbf{y}}) \times 10^{-2}$; Initial multiplier $\lambda_g^{(0)}=0, \forall g$; Available block index set $\mathcal{L} = \left\{g\right\}_{g=1}^G$}
\STATE {\textbf{\% ABSBL Core \%}}
\FOR {$t = 1:T_{ite}$}
\FOR {$l \in \mathcal{L}$}
\IF {mean$\left(\text{diag}(\mathbf{Q}_i) \right) <$ threshold}
\STATE {$\mathcal{L} = \mathcal{L} \setminus \{l\}$; Set $\boldsymbol{\mu}_\mathbf{x}^l = \mathbf{0}$, $\boldsymbol{\Sigma}_\mathbf{x}^l = \mathbf{0}_{UM \times UM}$;}
\ENDIF
\STATE Update $\gamma_{l,u}, \forall u$ by (\ref{absbl18});
\STATE Update $\bar{\mathbf{P}}$ by (\ref{absbl20}) and (\ref{absbl23});
\STATE Use ALM to update $\mathbf{P}_l$ and $\lambda_l$ by (\ref{absbl25});
\ENDFOR
\STATE {Update $\boldsymbol{\mu}_\mathbf{x}$ and $\boldsymbol{\Sigma}_\mathbf{x}$ by (\ref{absbl12});}
\STATE {Update $\sigma^2$ by (\ref{absbl22});}
\STATE {\textbf{if} $\|\boldsymbol{\mu}_\mathbf{x}^{(r+1)}-\boldsymbol{\mu}_\mathbf{x}^{(r)}\|_\text{F}^2 < \delta_1$ \textbf{then} \textbf{break};}
\ENDFOR
\STATE {\textbf{\% Off-grid Module \%}}
\FOR {$r = 1:R_{ite}$}
\STATE {Update $\mathbf{X}^{(r+1)}$ by (\ref{absbl27});}
\STATE {Compute $\nabla_{\mathbf{z}} f^{(r+1)}$ by (\ref{absbl28});}
\STATE {Update $\mathbf{z}^{r+1}$ by (\ref{absbl29});}
\STATE {\textbf{if} $\|\mathbf{X}^{(r+1)}-\mathbf{X}^{(r)}\|_\text{F}^2 < \delta_2$ \textbf{then} \textbf{break};}
\ENDFOR
\STATE {Rearrange to obtain $\hat{\mathbf{H}}$.}
\end{algorithmic}
\end{algorithm}

\subsection{Computational Complexity}
To compare the computational complexity of the algorithms, we tested with $N=GU$ antennas without subarray partitioning. The $\{\mathbf{Q}_g\}_{g=1}^G$ update costs $\mathcal{O}(G(U^3+MU^2))$. The $\{\mathbf{P}_g\}^{G}_{g=1}$ update, with $I$($<10$) ALM iterations and prior computation reuse, costs $\mathcal{O}(GIU^3)$ when using LU decomposition for determinant. Solving for $\boldsymbol{\mu}_\mathbf{x}$ and $\boldsymbol{\Sigma}_\mathbf{x}$ costs $\mathcal{O}(P N_{\text{RF}}N^2M^3  )$ utilizing block diagonal properties. Noise variance update costs $\mathcal{O}(M U G^2)$. Notably, ABSBL-MMV can reduce computational load by zeroing near-zero blocks after each iteration. The overall iteration complexity can be approximated as $O(T_\text{ite} (P N_{\text{RF}}N^2M^3+GU^3+GU^2 M+GIU^3+M U G^2))$.  Similarly, ABSBL's complexity is about $\mathcal{O}(T_\text{ite}M(GU^3+GIU^3+P N_{\text{RF}}N^2))$, where $T_\text{ite}$ represents iterations for the ABSBL algorithm. Our analysis shows that the MMV-assisted ABSBL algorithm trades computational efficiency for improved estimation accuracy. The off-grid module adds approximately $\mathcal{O}(R_\text{ite} P^2 N_\text{RF}^2 M+R_\text{ite} PN_\text{RF} M^2 )$ complexity, where $R_\text{ite}$ denotes iterations for off-grid module. {Hence, the proposed off-grid module can be turned on or off depending on deployment accuracy requirements and computational constraints.}

\subsection{Bayesian Cramér-Rao Bound}
The performance of inferred variables in Bayesian estimation is fundamentally limited by the BCRB \cite{5264921}. We analyze the BCRB to evaluate subarray-level channel recovery in multi-carrier systems.
\par We assume that the stacked sparse vector $\widetilde{\mathbf{x}}$ and the noise $\widetilde{\mathbf{n}}$ under multi-carrier are constrained by the distributions $\mathcal{C N}(\mathbf{0}, \hat{\mathbf{V}})$ and $\mathcal{C N}(\mathbf{0}, \hat{\sigma}_n^2\mathbf{I}_{MPN_{\text{RF}}})$, respectively, where $\hat{\mathbf{V}}=\operatorname{diag}\left\{\mathbf{I}_M \otimes \mathbf{\hat{Q}}_1 \mathbf{\hat{P}}_1 \mathbf{\hat{Q}}_1, \ldots, \mathbf{I}_M \otimes \mathbf{\hat{Q}}_G \mathbf{\hat{P}}_G \mathbf{\hat{Q}}_G\right\}$. The Bayesian Fisher information matrix (BFIM) of $\widetilde{\mathbf{x}}$ can be decomposed into a likelihood part and a prior part:
\begin{equation}
    \begin{aligned}
\mathbf{J}(\widetilde{\mathbf{x}})=-\mathbb{E}_{\widetilde{\mathbf{y}},\widetilde{\mathbf{x}}}\left[\frac{\partial^2 \ln p(\widetilde{\mathbf{y}}|\widetilde{\mathbf{x}})}{\partial \widetilde{\mathbf{x}} \partial \widetilde{\mathbf{x}}^\text{H}}\right]-
\mathbb{E}_{\widetilde{\mathbf{x}}}\left[\frac{\partial^2 \ln p(\widetilde{\mathbf{x}};\hat{\mathbf{V}})}{\partial \widetilde{\mathbf{x}} \partial \widetilde{\mathbf{x}}^\text{H}}\right]
\end{aligned}.
\end{equation}
\par Computing the Hessian matrix inside the expectation for both terms above yields: 
\begin{equation}
    \mathbf{J}(\widetilde{\mathbf{x}}) = \mathbb{E} \left[ \hat{\sigma}_n^{-2} \mathbf{I}_M \otimes(\boldsymbol{\Psi}^\text{H} \boldsymbol{\Psi})  + \hat{\mathbf{V}}^{-1} \right].
\end{equation}
\par The inverse of BFIM is the BCRB. Accordingly, the mean-squared error (MSE) of the $\widetilde{\mathbf{x}}$ is bounded by: 
\begin{equation}
\begin{aligned}
    \text{MSE}(\widetilde{\mathbf{x}})=&\mathbb{E}[\|\widetilde{\mathbf{x}}_{\text{est}}-\widetilde{\mathbf{x}}\|_2^2]\geq\text{tr}[\mathbf{J}^{-1}(\widetilde{\mathbf{x}})]\\=&M \ \text{tr}\left\{\left[ \hat{\sigma}_n^{-2} (\boldsymbol{\Psi}^\text{H} \boldsymbol{\Psi})  + \hat{\mathbf{V}}_\text{s}^{-1}\right]^{-1}\right\}, 
\end{aligned}
\end{equation}
where $\hat{\mathbf{V}}_\text{s}=\operatorname{diag}\left\{ \mathbf{\hat{Q}}_1 \mathbf{\hat{P}}_1 \mathbf{\hat{Q}}_1, \ldots,  \mathbf{\hat{Q}}_G \mathbf{\hat{P}}_G \mathbf{\hat{Q}}_G\right\}$.

\section{Simulation Results} \label{sec5}
\begin{figure*}[!t]
\centering
\includegraphics[width=6in]{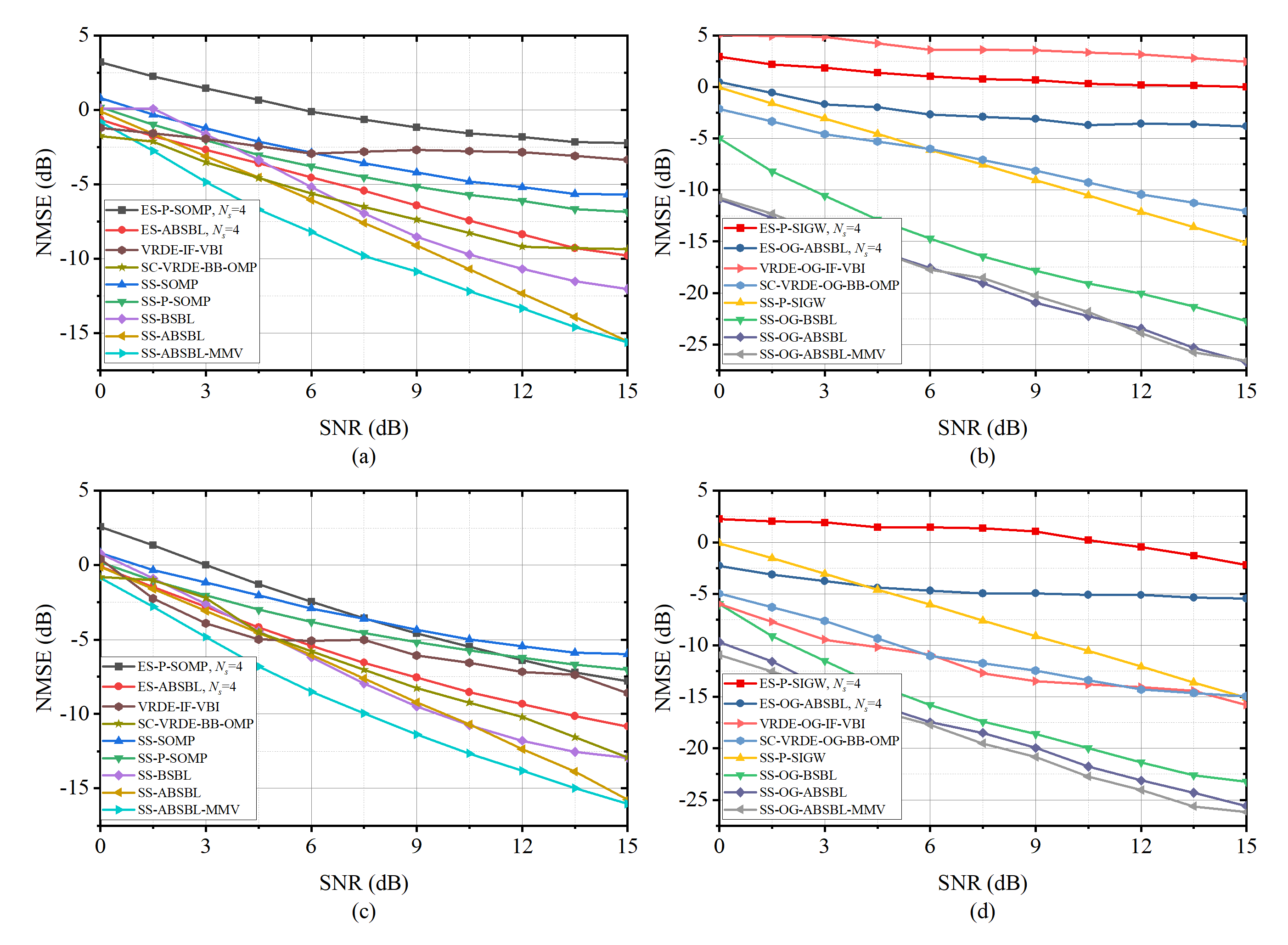}
\caption{NMSE versus SNR (a) $P=32$, On-grid, (b) $P=32$, Off-grid, (c) $P=64$, On-grid, (d) $P=64$, Off-grid.}
\label{fig_8}
\end{figure*}
In this section, we assess the performance of the proposed algorithm under various system settings. We use the normalized MSE as the metric: $\text{NMSE} = \mathbb{E} \left[\|\hat{\mathbf{H}} - \mathbf{H}\|^2_\text{F}/\|\mathbf{H}\|^2_\text{F} \right]$.

In this simulation, unless otherwise specified, we consider a multi-user XL-MIMO OFDM system. The BS is equipped with $N=512$ antenna elements and $N_\text{RF}=4$ RF chains, operating at a central frequency of $f_c=28$ GHz with a system bandwidth of  $B=100$ MHz. A total of $M=5$ subcarriers are simultaneously used to serve $K=3$ UEs. Each UE's channel has $L=3$ paths, with at most one being a non-ideal propagation path. The distances and angles of the scatters/UEs are sampled from $\mathcal{U}(10\text{m},100\text{m})$ and $\mathcal{U}(-2 \pi /3,2 \pi /3)$, respectively. The channels used in this simulation are generated by extending the code provided in \cite{10373799}, changing the uniform and fixed-position VRs distribution to a random size and position, and introducing possible non-ideal propagation with $t_d=1$. The minimum SI and sliding window size are set to 64. We perform 500 Monte Carlo simulations for statistical validation.

\subsection{Algorithm Description}
{To evaluate CE performance, we compare our proposed algorithm against several baselines, which can be grouped into three categories: methods using equal-sized subarray partitioning (ES- prefix), methods using the proposed adaptive segmentation (SS- prefix), and VR-prior methods. The baselines are summarized as follows:}

\begin{itemize}
  \item \textbf{ES-P-SOMP / ES-P-SIGW}\cite{10373799}: On-grid and off-grid versions of SOMP with a polar-domain codebook, using equal subarray partitioning and $N_s=4$ subarrays.
  \item \textbf{ES-ABSBL / ES-OG-ABSBL}: On-grid and off-grid versions of ABSBL with a DFT codebook, using equal partitioning and $N_s=4$ subarrays.
  \item \textbf{SS-(P-)SOMP / SS-P-SIGW}\cite{8306126}: On-grid and off-grid SOMP with either the polar-domain (P- prefix) or DFT codebook, employing the proposed PASS partitioning strategy.
  \item \textbf{SS-BSBL / SS-OG-BSBL}\cite{5887383}: On-grid and off-grid versions of BSBL with a DFT codebook, using the proposed PASS partitioning strategy.
  \item \textbf{SS-ABSBL(-MMV) / SS-OG-ABSBL(-MMV)}: On-grid and off-grid ABSBL with a DFT codebook, using PASS partitioning; the -MMV suffix denotes the MMV structured block-sparsity extension.
  \item \textbf{VRDE-IF-VBI / VRDE-OG-IF-VBI}\cite{10715712}: {A two-stage iterative framework. First, channel estimation is performed using an on-grid/off-grid version of IF-VBI with polar-domain codebook. Then, VR detection is carried out via structured expectation propagation. The two stages iterate until convergence.}
  \item \textbf{SC-VRDE-BB-OMP / SC-VRDE-OG-BB-OMP}\cite{10509715}: {Channels are first separated using MUSIC. For each subchannel, a two-stage procedure is applied: message passing for VR detection followed by an on-grid/off-grid version of BB-OMP with DFT codebook for channel estimation.}
  \end{itemize}

\subsection{Impact of SNR}
For fair comparison, on-grid and off-grid algorithms are evaluated separately. As shown in Fig. \ref{fig_8}(a)(c), for the on-grid architecture with equal subarray partitioning, the proposed ES-ABSBL algorithm outperforms ES-P-SOMP in terms of NMSE. Applying the proposed DHBF-PSSP improves the performance of P-SOMP and ABSBL, especially with low pilot length. Since the sparsity is not completely lost, the NMSE of SS-SOMP and SS-P-SOMP shows little difference. In contrast, the SS-BSBL and SS-ABSBL algorithms achieve decent performance without using high-complexity polar-domain codebooks. However, SS-BSBL underperforms SS-ABSBL at low pilot lengths due to its inability to simultaneously capture inter-block correlation and intra-block diversity. Furthermore, within the MMV framework, ABSBL fully utilizes the structured sparsity across subcarriers, delivering superior performance under low SNR. {It is worth noting that the two VR-prior methods, VRDE-IF-VBI and SC-VRDE-BB-OMP, suffer severe performance degradation under low pilot resources, because their VR-detection modules require message passing across the array dimension. When the number of observations is small, these modules cannot reliably recover accurate VR information.}
\par For off-grid algorithms, as shown in Fig. \ref{fig_8}(b)(d), all except ES-P-SIGW and ES-OG-ABSBL show significant performance gains, with the MMV framework having minimal impact. The performance loss of ES-P-SIGW and ES-OG-ABSBL is mainly due to non-ideal equal subarray partitioning, which leads to interpolation in the angular domain. This prevents on-grid algorithms from identifying the support set accurately, causing the off-grid module to reconstruct the channel matrix incorrectly or even fail to converge. The results once again demonstrate the significance of accurately capturing SnS.

\begin{figure}[!t]
  \centering
  \includegraphics[width=0.35\textwidth]{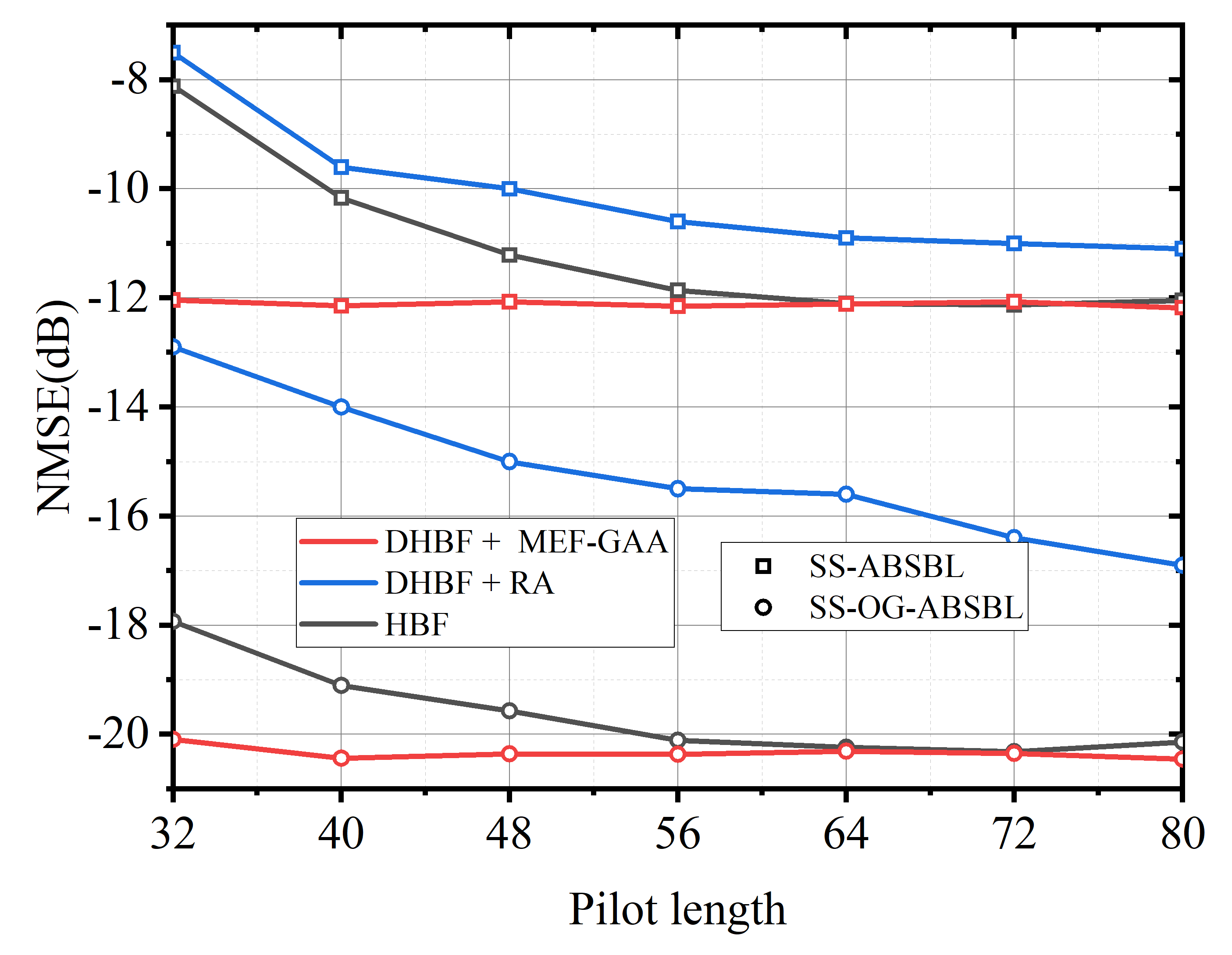}
  \caption{NMSE versus pilot length and beamforming architecture.}
  \label{fig_10}
\end{figure}

\subsection{Impact of Pilot Length and Beamforming Architecture}
\par {To verify the advantage of the DHBF-PSSP architecture in reducing pilot overhead for CE, we evaluate three architectures: fully connected HBF, DHBF with random RF-chain allocation (RA), and DHBF assisted by MEF-GAA (i.e., the proposed DHBF-PSSP) with ABSBL and OG-ABSBL, under an SNR of 10 dB.}
\par As shown in Fig. \ref{fig_10}, two algorithms exhibit significant performance gains with the DHBF-PSSP architecture using the MEF-GAA algorithm, especially at low pilot lengths. For a given number of pilots and known subarray partitioning, the DHBF architecture with MEF-GAA effectively increased the number of decoupled effective pilots per subarray, confirming DHBF-PSSP's enhanced capability in optimizing pilot allocation and improving CE accuracy. {However, without a proper RF chain allocation strategy, DHBF may perform even worse than the fully connected HBF architecture.}

\begin{figure}[!t]
  \centering
  \includegraphics[width=0.35\textwidth]{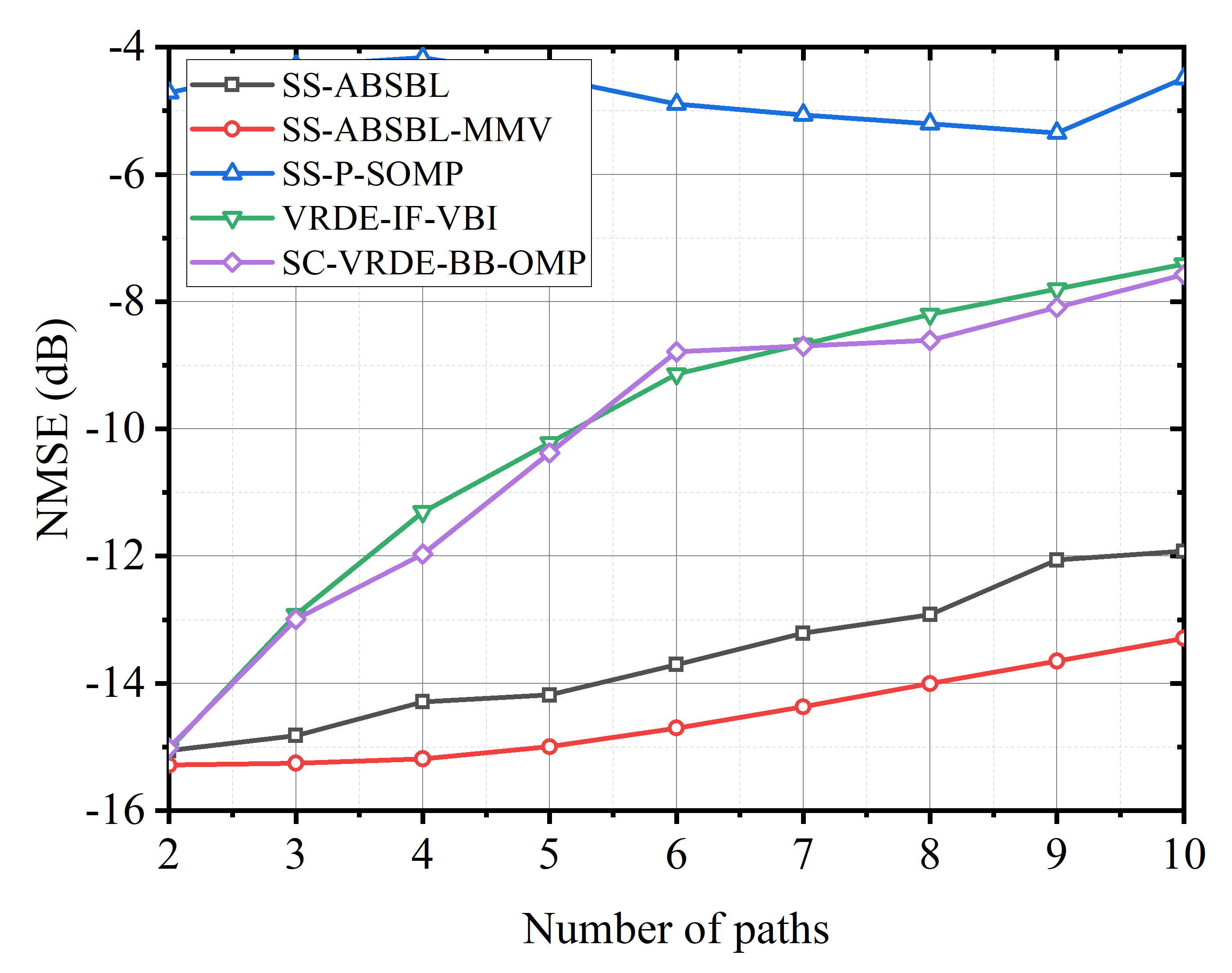}
  \caption{{NMSE versus the number of paths.}}
  \label{com_path}
\end{figure}

\subsection{{Impact of the number of paths}}
{In Fig. \ref{com_path}, we evaluate the impact of the number of channel paths on algorithm performance at SNR = 15 dB. The path count is varied from 2 to 10, and the pilot length is set to $P=64$. As the channel becomes less sparse, all methods exhibit performance degradation. Nevertheless, the proposed algorithm consistently outperforms the baselines. This robustness is attributable to the SBL framework’s soft-sparsity mechanism. By contrast, VR-prior methods deteriorate because their VR-detection modules must cope with an increasing number of paths. And sub-channel approaches suffer from increased inter-subchannel interference as the path count grows.}

\begin{figure}[!t]
  \centering
  \includegraphics[width=0.35\textwidth]{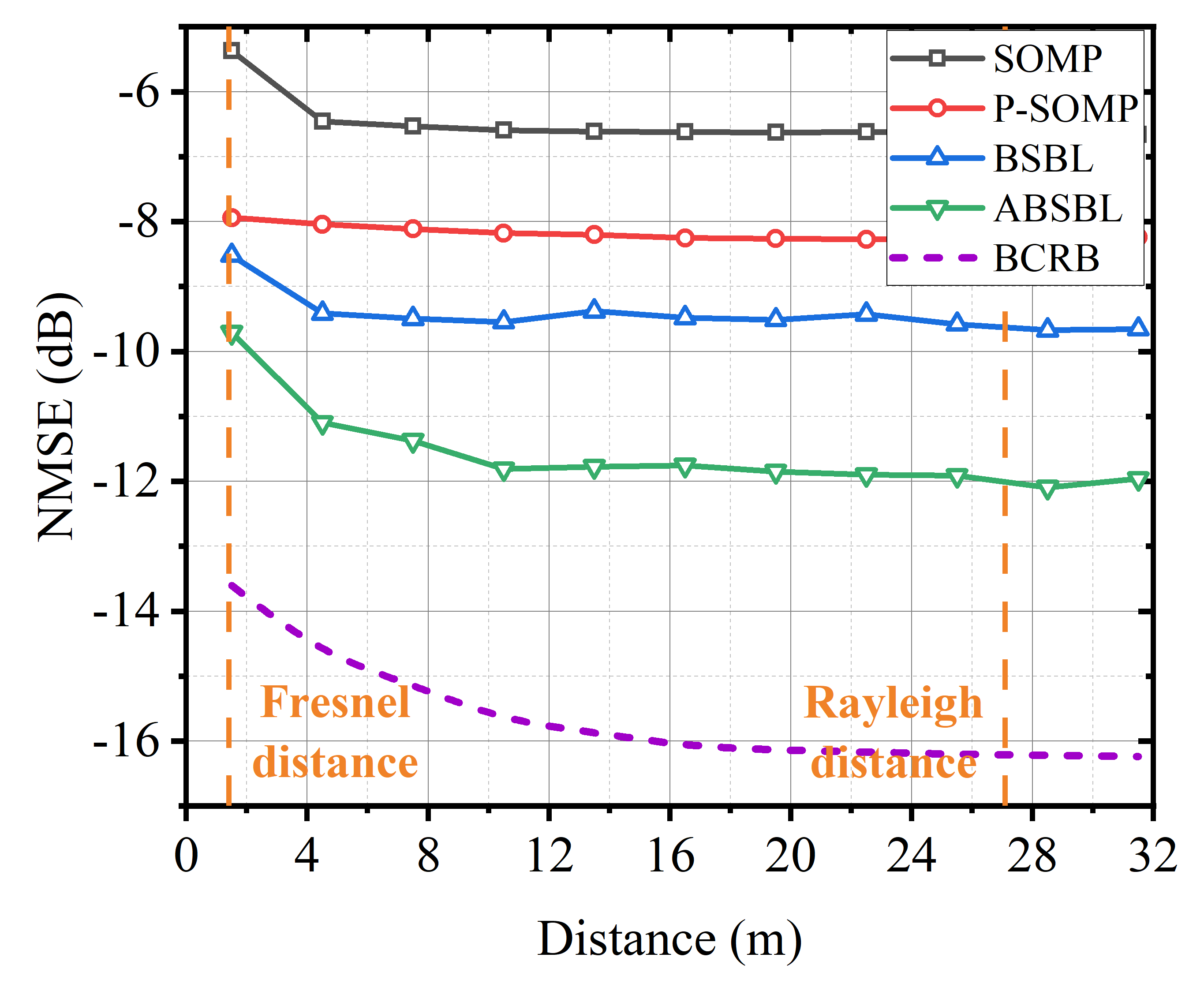}
  \caption{NMSE versus UE-BS distance.}
  \label{fig_11}
\end{figure}

\subsection{Impact of Distance}
Transforming the near-field spatial channel to the angular domain reveals that decreased distance leads to a significant increase in energy leakage, consequently reducing angular domain sparsity. This sparsity reduction directly degrades the performance of CE algorithms relying on CS techniques. To specifically assess the proposed algorithm's performance under different sparsity levels, this simulation concentrates on the impact of UE-BS distance, excluding the effects of SnS. The simulation setup employs a fully connected HBF architecture with $N=128$ antennas and $N_\text{RF}=4$ RF chains. The pilot length and SNR are set to $P=20$ and 10 dB. The UE-BS distance gradually increases from 1.5 m to 31.5 m. In this system, the Fresnel distance is $R_f = 0.62 \sqrt{\frac{[(N-1)d]^3}{\lambda}} \approx 1.41 \, \text{m}$, and the Rayleigh distance is $R_r = \frac{2[(N-1)d]^2}{\lambda} \approx 27.10 \, \text{m}$. The distance range set in the simulation covers both the near-field and far-field regions. As SnS is not taken into account, the BCRB serves as the NMSE performance bound in this simulation.
\par As depicted in Fig. \ref{fig_11}, the performance of all algorithms decreases when the UE-BS distance is less than 10 meters. This decline is attributed to the loss of sparsity due to angular spreading for DFT-based methods and insufficient resolution for P-SOMP due to fewer distance samples. Our proposed ABSBL algorithms consistently outperform others across different distances. Algorithm performance notably stabilizes around 10 meters, significantly less than the conventional Rayleigh distance.  This indicates that the Rayleigh distance might overestimate the detrimental impacts of the near-field on CE. Thus, recent research has introduced the concept of the effective Rayleigh distance to provide a more precise definition of strong and weak near-field channels \cite{10541333,9617121}.

\begin{figure}[!t]
  \centering
  \includegraphics[width=0.35\textwidth]{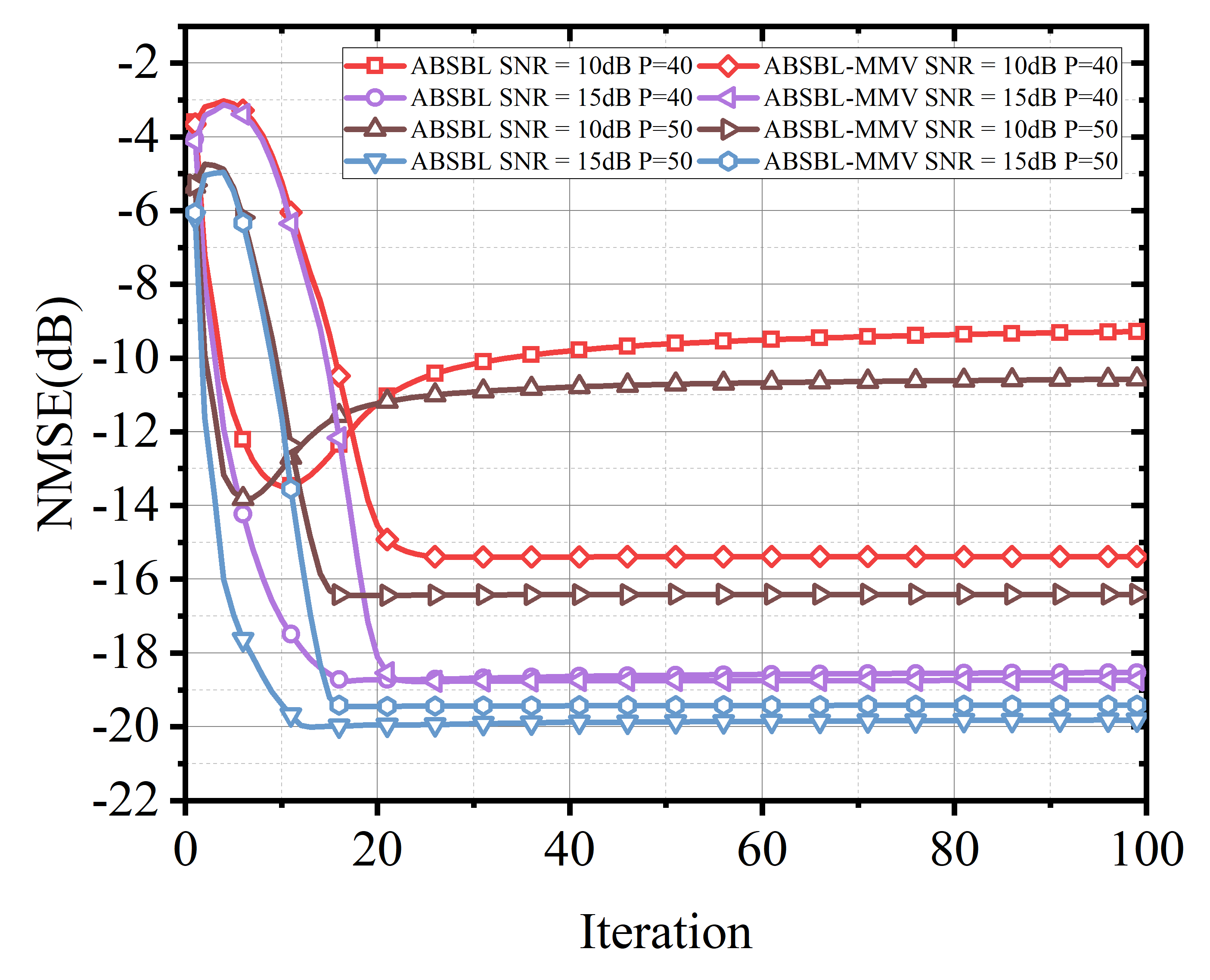}
  \caption{The convergence of ABSBL.}
  \label{fig_12}
\end{figure}

\subsection{Convergence}
To assess the convergence performance of the proposed ABSBL and ABSBL-MMV algorithms, we conduct systematic simulations with the following setup: $N=256$, $N_\text{RF}=4$, and the pilot length is set to $P=40$ and $P=50$. The SNR is set at two levels, 10 dB and 15 dB. For each parameter combination, we perform 100 iterations to observe the change in NMSE over iterations.
\par As depicted in Fig. \ref{fig_12}, two key observations emerge. First, across all tested configurations, NMSE decreases with increasing iterations, indicating good convergence. Second, at low SNR (10 dB), the ABSBL shows a characteristic convergence curve that initially decreases, then increases, and finally stabilizes, which is not observed at high SNR (15 dB). This error change stems from the sparsity induction mechanism of ABSBL: precision parameter adjustment initially prioritizes critical feature retention, but prolonged iterations may retain noise-corrupted or redundant features, causing minor error resurgence. Notably, the ABSBL-MMV, which leverages inter-carrier structured sparsity, demonstrates significant performance advantages in low SNR environments, with faster and more stable convergence than the ABSBL.
\par Comparing results across parameter settings shows that SNR affects algorithm performance more significantly than pilot length. Additionally, Fig. \ref{fig_12} serves as a crucial guide for selecting the required CE iterations. Our analysis suggests that 30 iterations achieve accurate and stable CE within our simulation setup.

\section{Conclusion and Future work} \label{sec6}
This paper explored the subarray-based near-field CE problem under SnS effects. We first established the SnS near-field channel model for XL-MIMO systems. Given that existing SnS near-field CE algorithms assumed equal array partitioning, we analyzed the limitations of non-ideal subarray segmentation. To address this, we proposed a DHBF-PSSP architecture, which included a PASS algorithm for measurement-driven array partitioning, an MEF-GAA for RF chain resource allocation under limited RF chain constraints, and an SS-SM for subarray decoupling via DHBF. We then developed both on-grid and off-grid versions of the SS-ABSBL-MMV algorithm for subarray CE, which employed the block sparsity in angular-domain channels and structured sparsity across subcarriers, thereby reducing computational and storage overhead. Simulation results demonstrated that the proposed framework achieved superior NMSE performance. 
\par {Research on near-field channels is still at an early stage. Their characteristics and the corresponding signal processing methods need further investigation. Future work may focus on: 1) the conditions under which SnS occurs and whether SnS can serve as a user fingerprint to improve communication performance; 2) RF chain resource optimization for receive-side DHBF architectures; 3) unified far- and near-field channel representations and associated signal processing methods, e.g., the wavenumber domain; 4) more effective definition of the near- and far-field boundary; 5) dynamic and joint-users near-field channel estimation.}

\footnotesize
\bibliographystyle{IEEEtran}
\bibliography{IEEEabrv,IEEEexample}

\vfill

\end{document}